\documentclass[fleqn,usenatbib]{mnras}
\usepackage{newtxtext,newtxmath}

\usepackage[T1]{fontenc}
\usepackage{longtable}
\usepackage{soul}
\usepackage{multirow}
\usepackage{pdflscape}

\DeclareRobustCommand{\VAN}[3]{#2}
\let\VANthebibliography\thebibliography
\def\thebibliography{\DeclareRobustCommand{\VAN}[3]{##3}\VANthebibliography}

\usepackage{graphicx}
\usepackage{amsmath}
\usepackage[LGRgreek]{mathastext}
\usepackage{array}
\usepackage{mathtools}
\usepackage{booktabs}
\usepackage{float}
\usepackage{subcaption}
\usepackage[english]{babel}
\usepackage{hyperref}
\addto\extrasenglish{
    
}
\usepackage{longtable}

\title[TRAPUM Large Magellanic Cloud survey II]{The TRAPUM Large Magellanic Cloud pulsar survey with MeerKAT II: 12 new discoveries and timing solutions for 7 pulsars}

\author[V. Prayag et al.]{\parbox{\textwidth}{
V. Prayag,$^{1,2}$\thanks{E-mail: \href{mailto:venu.prayag@gmail.com}{venuprayag@gmail.com}}
L. Levin,$^{3}$
M. Geyer, $^{2,1}$
B.~W.~Stappers,$^{3}$
H. Hurter,$^{4}$
E.~D.~Barr,$^{5}$
S. Buchner,$^{6}$
M. Burgay,$^{7}$
F. Calore,$^{8}$
E. Carli,$^{9,10}$
M. Colom i Bernadich,$^{5,11}$
L. Gebauer-Werner,$^{5}$
M. Kramer,$^{5}$
P.~V.~Padmanabh,$^{12,13}$
A. Ridolfi,$^{14}$
T.~Thongmeearkom,$^{15,3}$
J. D. Turner,$^{3}$
C. Venter$^{4,16}$
}
\\ \\ \\
$^{1}$Department of Astronomy, The University of Cape Town, Private Bag X3, Rondebosch 7701, Cape Town, South Africa\\
$^{2}$High Energy Physics, Cosmology \& Astrophysics Theory (HEPCAT) Group, Department of Mathematics \& Applied Mathematics, \\University of Cape Town, Cape Town 7700, South Africa \\
$^{3}$Jodrell Bank Centre for Astrophysics, Department of Physics and Astronomy, The University of Manchester, Manchester M13 9PL, United Kingdom \\
$^{4}$Centre for Space Research, North-West University, Private Bag X6001, Potchefstroom 2520, South Africa\\
$^{5}$Max-Planck-Institut für Radioastronomie, Auf dem Hügel 69, D-53121 Bonn, Germany\\
$^{6}$South African Radio Astronomy Observatory (SARAO), 2 Fir Street, Black River Park, Observatory, Cape Town, 7925 \\
$^{7}$INAF -- Osservatorio Astronomico di Cagliari, via della Scienza 5, 09047 Selargius (CA), Italy \\
$^{8}$ LAPTh, CNRS, USMB, F-74940 Annecy, France\\
$^{9}$Centre for Astrophysics and Supercomputing, Swinburne University of Technology, Hawthorn VIC 3122, Australia
\\
$^{10}$OzGrav: The ARC Center of Excellence for Gravitational Wave Discovery, Hawthorn VIC 3122, Australia\\
$^{11}$INAF -- Osservatorio Astronomico di Cagliari, via della Scienza 5, 09047 Selargius (CA), Italy \\
$^{12}$Max Planck Institute for Gravitational Physics (Albert Einstein Institute), D-30167 Hannover, Germany\\
$^{13}$Leibniz Universit{\"a}t Hannover, D-30167 Hannover, Germany\\
$^{14}$ Fakult\"at f\"ur Physik, Universit\"at Bielefeld, Postfach 100131, D-33501 Bielefeld, Germany\\
$^{15}$National Astronomical Research Institute of Thailand, Don Kaeo, Mae Rim, Chiang Mai 50180, Thailand \\
$^{16}$National Institute for Theoretical and Computational Sciences, South Africa \\
}
\date{Accepted XXX. Received YYY; in original form ZZZ}

\pubyear{2025}

\begin{document}
\label{firstpage}
\pagerange{\pageref{firstpage}--\pageref{lastpage}}
\maketitle

\begin{abstract}
We report the discovery of 12 new radio pulsars in the Large Magellanic Cloud (LMC) as part of the TRAPUM (TRAnsients and PUlsars with MeerKAT) Large Survey Project, using the MeerKAT L-band receivers (856--1712\,MHz). These pulsars, discovered in 18 new pointings with 2\,hour integration times, bring the total number of pulsars identified by this ongoing survey to 19 (yielding a total of 44 LMC radio pulsars now known), representing an 80 per cent increase in the LMC radio pulsar population to date. These include PSR\,J0454$-$6927, the slowest extragalactic radio pulsar discovered to date, with a spin period of 2238\,ms, and PSR\,J0452$-$6921, which exhibits the highest dispersion measure (DM) among extragalactic radio pulsars, at 326\,pc\,cm$^{-3}$. The fastest spin period among the new discoveries is 245\,ms, and the lowest DM is 62\,pc\,cm$^{-3}$. We also present timing results for our first pulsar discoveries with MeerKAT and the Murriyang radio telescope, obtaining phase-connected solutions for seven pulsars in the LMC. These results indicate that the pulsars are isolated, canonical radio pulsars with characteristic ages up to 8.8\,Myr.
\end{abstract}


\section{Introduction}

The Magellanic Clouds, comprising the Large Magellanic Cloud (LMC) and the Small Magellanic Cloud (SMC), are ideal targets for radio pulsar searches due to their proximity to the Milky Way, high star formation rates, and environmental properties contrasting with those of our galaxy, such as lower metallicity. The SMC and LMC, at distances of 62.4\,kpc and 49.6\,kpc respectively \citep{Graczyk2020a,Graczyk2020b}, host the only radio pulsars discovered outside the Milky Way to date, with 14 in the former \citep{McConnell1991, Crawford2001, Manchester2006, Titus2019, Carli2024a} and 32 in the latter, including seven discovered in \citet{Prayag2024}, Paper\,I of this series \citep{McCulloch1983, Seward1984, McConnell1991, Crawford2001, Manchester2006, Ridley2013, Hisano2022, Wang2022, Xia2025}.

TRAPUM (Transients and Pulsars with MeerKAT)\footnote{\href{http://www.trapum.org/}{http://www.trapum.org/}} is a large survey project focused on finding new pulsars and exploring transient phenomena in the radio band \citep{Stappers2016} using the 64-dish MeerKAT radio telescope array \citep{Jonas2016, Camilo2018}, located in South Africa. Amongst the many objectives of TRAPUM is the survey of the Magellanic Clouds. In Paper\,I, we reported the first seven radio pulsar discoveries from the ongoing TRAPUM LMC Survey, with dispersion measures (DMs) ranging from 79 to 254\,$pc \, cm^{-3}$ and periods ranging from 278 to 1690\,ms. These pulsars do not appear to be associated with any known sources, such as supernova remnants (SNRs) or pulsar wind nebulae (PWNe). We subsequently characterise some of these discoveries in this paper using pulsar timing.

Pulsar timing models times of arrival (ToAs) of the pulses with the rotational and orbital parameters of a pulsar. This process involves fitting the pulsar’s spin-down, positional, DM, and possibly binary parameters to the measured ToAs across observing epochs, refining the fit until the differences from the timing model are minimised and the number of rotations between pulse arrivals is unambiguously determined. Determining these primary parameters through timing allows us to calculate derived properties such as the characteristic age, spin-down luminosity, and surface magnetic field strength of the pulsar \citep[see][]{Lorimer2004}. In this paper, we present results from 18 new search pointings of the ongoing TRAPUM LMC survey that have resulted in the discovery of 12 new pulsars, bringing the number of pulsars discovered by the survey to 19, and we provide timing solutions for seven of these sources.

The paper is structured as follows: in \autoref{section: observations}, we provide an overview of the search observations and describe our approach to performing the timing observations. We outline how the processing of the search and timing analysis was carried out in \autoref{section: processing}. The results from the 18 new pointings of the TRAPUM LMC Survey and the timing campaigns are presented in \autoref{section: results}. In \autoref{section: discussion and conclusions}, we discuss and conclude on our findings.

\section{Observations} \label{section: observations}
\subsection{Search observations}

A total of 18 search pointings, each 2\,h in duration, in addition to the four reported in Paper\,I, were conducted between 2023 June and 2024 November using the MeerKAT L-band receivers, covering 856--1712\,MHz with a central frequency of 1284\,MHz. A summary of the observation parameters is given in \autoref{Appendix A: observations} of the Appendix. The data were recorded using 2048 frequency channels and a sampling time of 153\,$\upmu s$. Using MeerKAT’s core dishes (44 antennas within a 1\,km radius) for efficient sky coverage and sensitivity, this configuration enabled the Filterbank and Beamforming User Supplied Equipment (FBFUSE; \citealt{Barr2018}), supported by the Accelerated Pulsar Search User Supplied Equipment (APSUSE; \citealt{Padmanabh2023} and references therein), to generate up to 768 tied-array coherent beams (CBs) as part of the TRAPUM backend. We targeted known sources that could be associated with pulsars, such as SNRs, candidate SNRs, globular clusters (GCs), and PWNe (see \autoref{appendix: Targeted sources} of the Appendix for a full list of sources present in our pointings). Pointing\,7 was re-observed as Pointing\,12 due to unexpected data recording issues, with the centre of Pointing\,12 slightly offset from that of Pointing\,7. The beam tiling maps for each of the 18 pointings are shown in \autoref{appendix: beam maps} of the Appendix. We follow the same search setup as described in Paper\,I, where the details of the observations' configuration can be found.

\begin{center}
\begin{figure*}
\begin{tabular}{c}
\includegraphics[width=2\columnwidth]{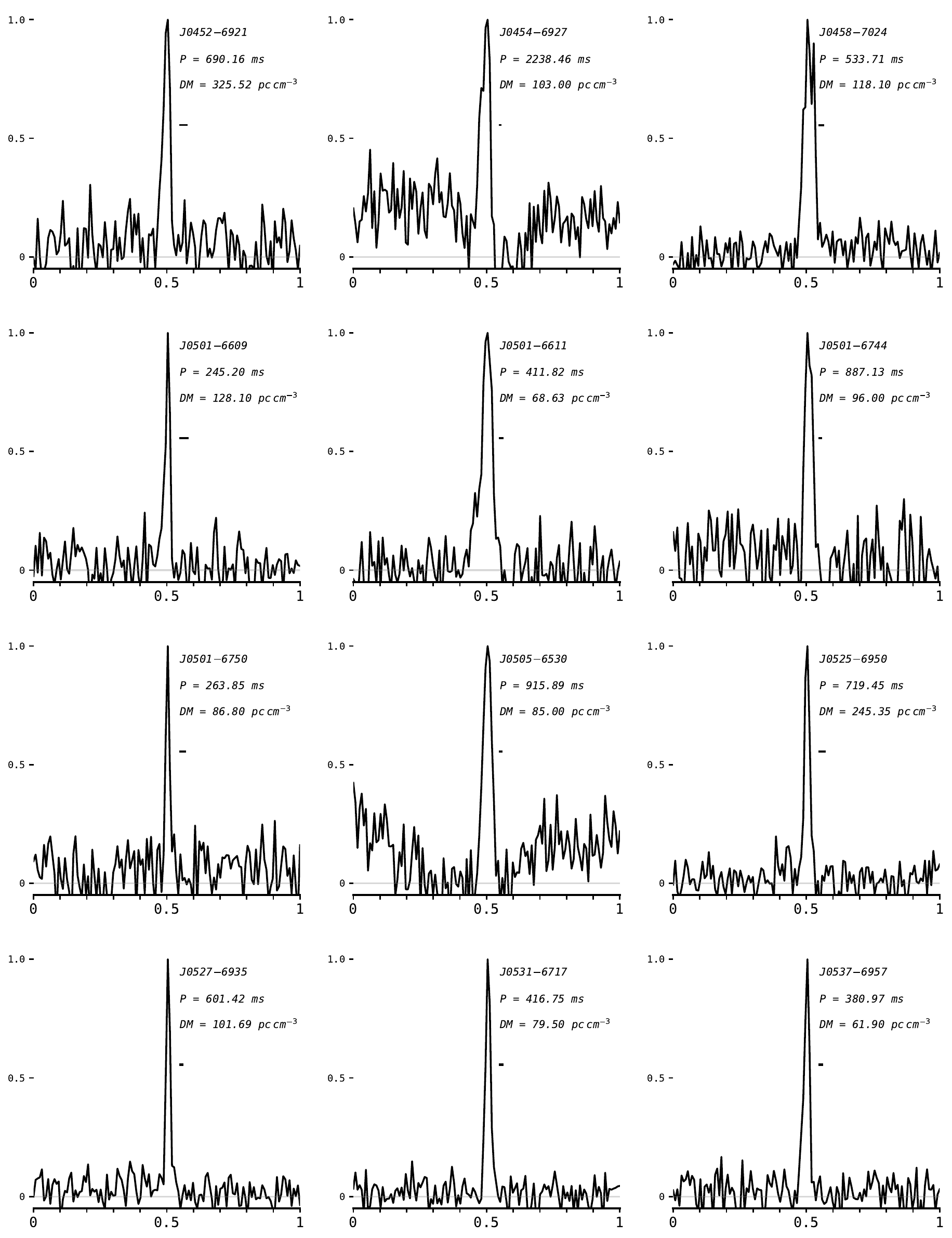} \\
\end{tabular}
\caption{The integrated L-band discovery pulse profiles obtained using our search pipeline (see \autoref{section: Search}), along with the periods and DMs, are shown for the newly discovered pulsars. The derived pulsar properties obtained from the search pipeline are listed in \autoref{table: Discoveries}. Each plot represents one full rotation of the pulsar with the x-axis showing the rotational phase, 0 to 1, divided into 128 phase bins, and the y-axis representing the pulse amplitude, normalised so that each profile has a unit peak. The grey horizontal line marks the zero level, and the width of the black horizontal bar indicates the effective time resolution of the system relative to each pulsar’s spin period.}
\label{figure: pulse profiles}
\end{figure*}
\end{center}

\subsection{Timing observations} \label{section: Timing observations}

Following the discoveries, we initiated pulsar timing observations with MeerKAT at pseudo-logarithmic intervals in time at L- and UHF-band (544--1088\,MHz), using the full array of available dishes, spanning a few months. Short observing intervals allow us to efficiently capture and characterise rapid variations in the pulsar’s period, while observations spaced over days to weeks provide sensitivity to longer-term trends. This strategy ensures that phase connection is maintained across both short- and long-term timescales. As such, we typically have two observations on the first day, Day\,1, separated by a few hours to facilitate phase connection, followed by observations on Day\,2, Day\,5, Day\,10, Day\,20, Day\,50, and Day\,100. Similar to the search observations, we used FBFUSE to digitally beamform CBs with 70 per cent overlap at the pulsar locations obtained using the \textsc{seeKAT} Python package\footnote{\href{https://github.com/BezuidenhoutMC/SeeKAT}{https://github.com/BezuidenhoutMC/SeeKAT} accessed on 2025 April 10} \citep{Bezuidenhout2023} from the discovery data, where applicable, as described in Paper\,I. The closer CB spacing reduced sensitivity loss between beams and improved positional accuracy. Once we obtained at least three CB detections from a timing observation of a pulsar, we used \textsc{seeKAT} to improve the localisation and centered the next observation on the updated position. Iterating this process allowed us to achieve a better signal-to-noise ratio (S/N) and further refine each pulsar’s position before re-observing, enabling us to use only one CB per pulsar once an adequate level of precision had been achieved. Moreover, the better S/N allowed for a reduction in observation length in some cases.

The first timing campaign involved pseudo-logarithmically spaced MeerKAT L-band observations of some of the discoveries reported in Paper\,I, namely PSRs\,J0509$-$6838, J0509$-$6845, J0518$-$6939, and J0519$-$6931. PSRs\,J0518$-$6946, J0534$-$6905, and the pulsar discovered only in the incoherent beam of Pointing\,1 (named IB pulsar) were not included in the first campaign. This was because the IB pulsar was not well localised, the timing observation coherent beam covering PSR\,J0518$-$6946 yielded only weak or no detections, and PSR\,J0534$-$6905 was discovered only after the campaign had already begun. Ten observations were obtained for each pulsar over approximately 260 days, from 2022 December to 2023 August. The L-band observations used a bandwidth of 856\,MHz split into 4096 frequency channels, with a sampling time of 76\,$\upmu$s, and were recorded in filterbank format on the APSUSE cluster. 

The second timing campaign involved MeerKAT UHF-band observations, with a bandwidth of 544\,MHz, a sampling time of 120\,$\upmu$s, and 4096 frequency channels. The campaign targeted PSRs\,J0501$-$6609, J0501$-$6611, and J0531$-$6717, which are reported as discoveries in this paper, as well as PSR\,J0534$-$6905. The remaining new pulsars reported in this paper were discovered after the second timing campaign had already begun. The switch to the UHF band is motivated by the typically negative spectral indices of pulsars \citep{Jankowski2018}, making them brighter at lower frequencies and allowing for shorter observations. These observations, totalling up to 11 epochs, spanned approximately 130 days, from 2024 June to 2024 November.

Most of the pulsars discovered so far are not bright enough to be timed using the 64\,m Parkes ``Murriyang'' radio telescope, located in New South Wales, Australia. However, for the three brightest discoveries from the survey up to now, PSRs\,J0509$-$6845, J0501$-$6611, and J0531$-$6717, we used Murriyang to obtain additional timing observations\footnote{This was done under the P1054 project, `Follow-up of pulsar discoveries from MeerKAT searches'}. We used the ultra-wide-bandwidth low-frequency receiver (UWL, \citealt{Hobbs2020}), operating from 704 to 4032\,MHz in search mode, with a sampling time of 64\,$\upmu$s and 3328 frequency channels ($26\,subbands\times128\,channels$).

\section{Processing} \label{section: processing}

\subsection{Search} \label{section: Search}

The TRAPUM pulsar search pipeline consists of three main stages: searching, filtering, and folding. During the search stage, radio frequency interference (RFI) removal was performed using \texttt{filtool} from the \textsc{pulsarX}\footnote{\href{https://github.com/ypmen/PulsarX}{https://github.com/ypmen/PulsarX} accessed on 2025 April 03} suite \citep{men2023}, and a de-dispersion plan covering DMs of 35--500\,$pc\,cm^{-3}$ was implemented. A GPU-accelerated Fourier-domain periodicity search was performed using \textsc{peasoup}\footnote{\href{https://github.com/ewanbarr/peasoup}{https://github.com/ewanbarr/peasoup} accessed on 2025 April 03} \citep{Barr2020}, targeting periods up to 10\,s. Both full 2\,h searches and 20-minute segment searches were conducted, the latter to increase sensitivity to pulsars in binaries with shorter orbital periods. Acceleration searches were not performed in order to increase processing efficiency. Multi-beam filtering further mitigates RFI, and candidates were grouped based on spin period, DM, beam location, and, where applicable, acceleration values. In the folding stage, \textsc{pulsarX}’s \texttt{psrfold\_fil} was used to fold only those candidates with peak S/N values above 9.5, a threshold chosen to balance processing speed with sensitivity. The resulting folded candidates were then classified using the Pulsar Image-based Classification System (\textsc{pics}, \citealt{Zhu2014}), a deep learning model for image-based pattern recognition in pulsar identification, trained on survey data including that from the TRAPUM project. High-ranking candidates were inspected, and from these, new pulsar discoveries were identified. The surrounding beams were refolded using the candidate parameters to search for additional detections, which were then used to improve the localisation of each source with \textsc{seeKAT}. A comprehensive overview of the search pipeline is provided in \cite{Padmanabh2023}, \cite{Carli2024a}, and Paper\,I.

\subsection{Timing}

We used the DM and period of each pulsar, obtained from their respective search discovery beams, to initially fold the MeerKAT timing observation data using \texttt{psrfold\_fil}, generating one \textsc{psrchive}-format \citep{Hotan2004, VanStraten2012} folded data archive file for each CB, after which we identified the beam with the highest S/N. The \texttt{filtool}-cleaned filterbank data from this beam was then folded with \textsc{dspsr} \citep{vanStraten2011} using a parameter file with the best DM, period, and localisation (see \autoref{section: Timing observations}), producing a high-resolution archive with 10-second sub-integrations, 4096 frequency channels, and 2048 phase bins per rotation. The high-resolution archive files were further cleaned of RFI using \textsc{clfd} \citep{Morello2023}, with additional manual cleaning, where required, performed using \textsc{psrchive}’s \texttt{pazi}.

Using \textsc{psrchive}'s \texttt{psradd}, we combined the high-resolution archive files to make a single phase-aligned pulse profile and subsequently created a noiseless template using \textsc{psrchive}'s \texttt{paas}. \textsc{psrchive}'s \texttt{pat} was used to compute ToAs and their associated uncertainties by cross-correlating the observed pulse profile with the noiseless template. We then fitted a pulsar timing model to the ToAs with \textsc{tempo2} \citep{Hobbs2006, Edwards2006}, adjusting spin, position, orbital elements, and DM where necessary, while deterministic clock and ephemeris corrections were applied automatically and instrumental jumps included only when required. We used the DE405 Solar System ephemeris from the Jet Propulsion Laboratory (JPL), available in \textsc{tempo2} version 2023.05.1, to model planetary motions and other Solar System dynamics when converting topocentric ToAs to barycentric ToAs.

Regarding the Murriyang observations, we first used \texttt{digifil} from \textsc{dspsr} to convert the search-mode data into single filterbank files, which were then processed as described for the MeerKAT timing data. For each pulsar, the Murriyang noiseless template was rotated to match the phase of the corresponding MeerKAT template, applying a \texttt{JUMP} parameter only if necessary.

\begin{figure}
\begin{center}
\includegraphics[width=0.7\columnwidth]{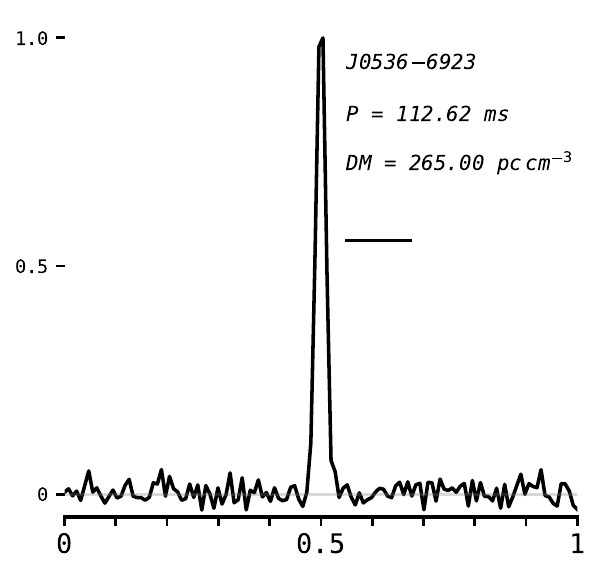}
\caption{The integrated pulse profile of our first observation of PSR\,J0536$-$6923. This pulsar was discovered by \protect\cite{Ridley2013} in Murriyang data and localised in our observations (see \autoref{section: Localisation of PSR J0537-69}). Remaining details of the plot are as described in the caption of \autoref{figure: pulse profiles}.}
\label{figure: 0537 pulse profile}
\end{center}
\end{figure}

\section{Results} \label{section: results}

\begin{figure}
\includegraphics[width=\columnwidth]{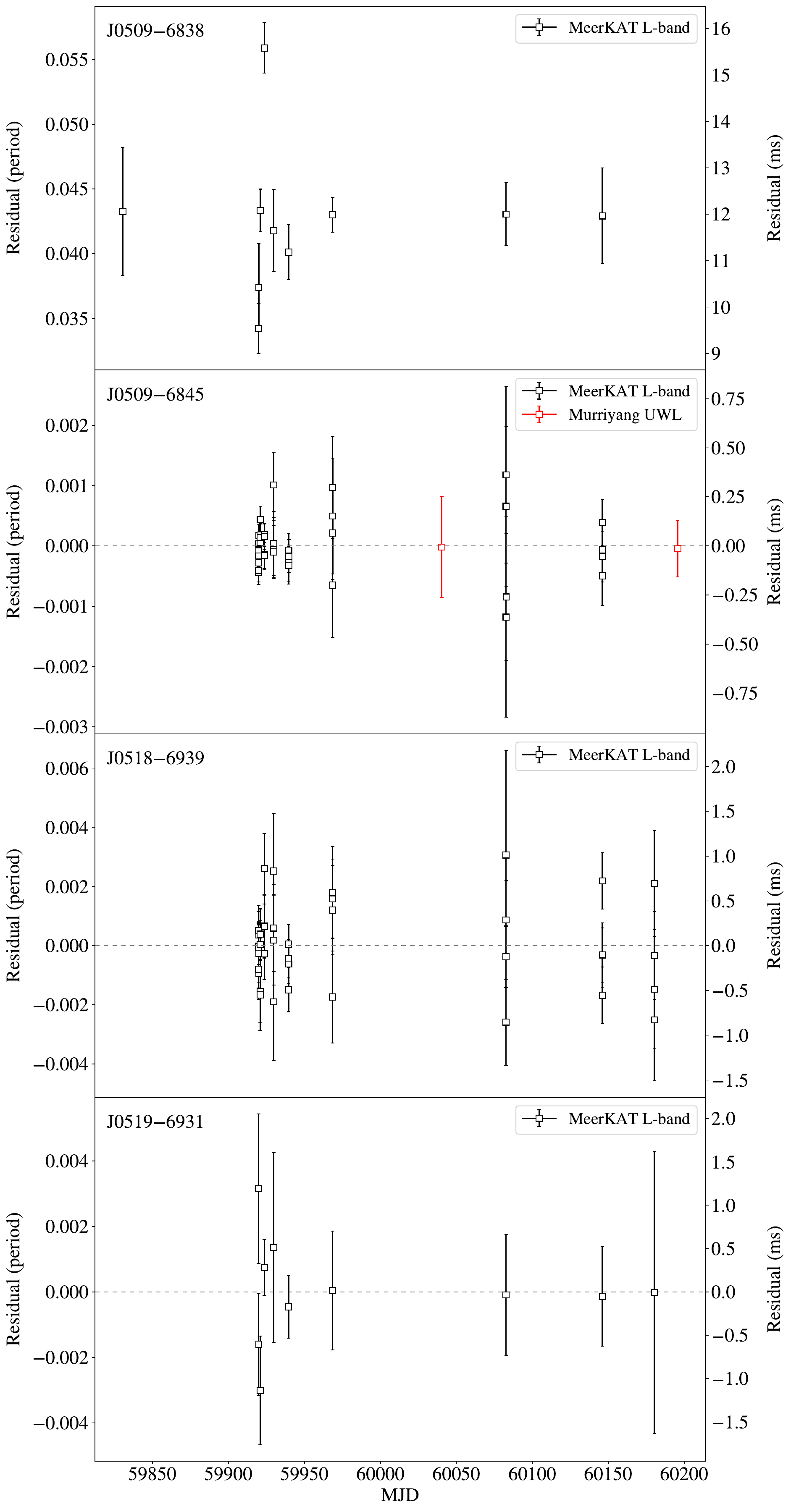}
\caption{The Tempo2 timing residuals for PSRs\,J0509$-$6838, J0509$-$6845,
J0518$-$6939, and J0519$-$6931 (refer to \autoref{table: L-band campaign}). These pulsars were part of the first timing campaign.}
\label{figure: timing residuals, first campaign}
\end{figure}

\begin{figure}
\includegraphics[width=\columnwidth]{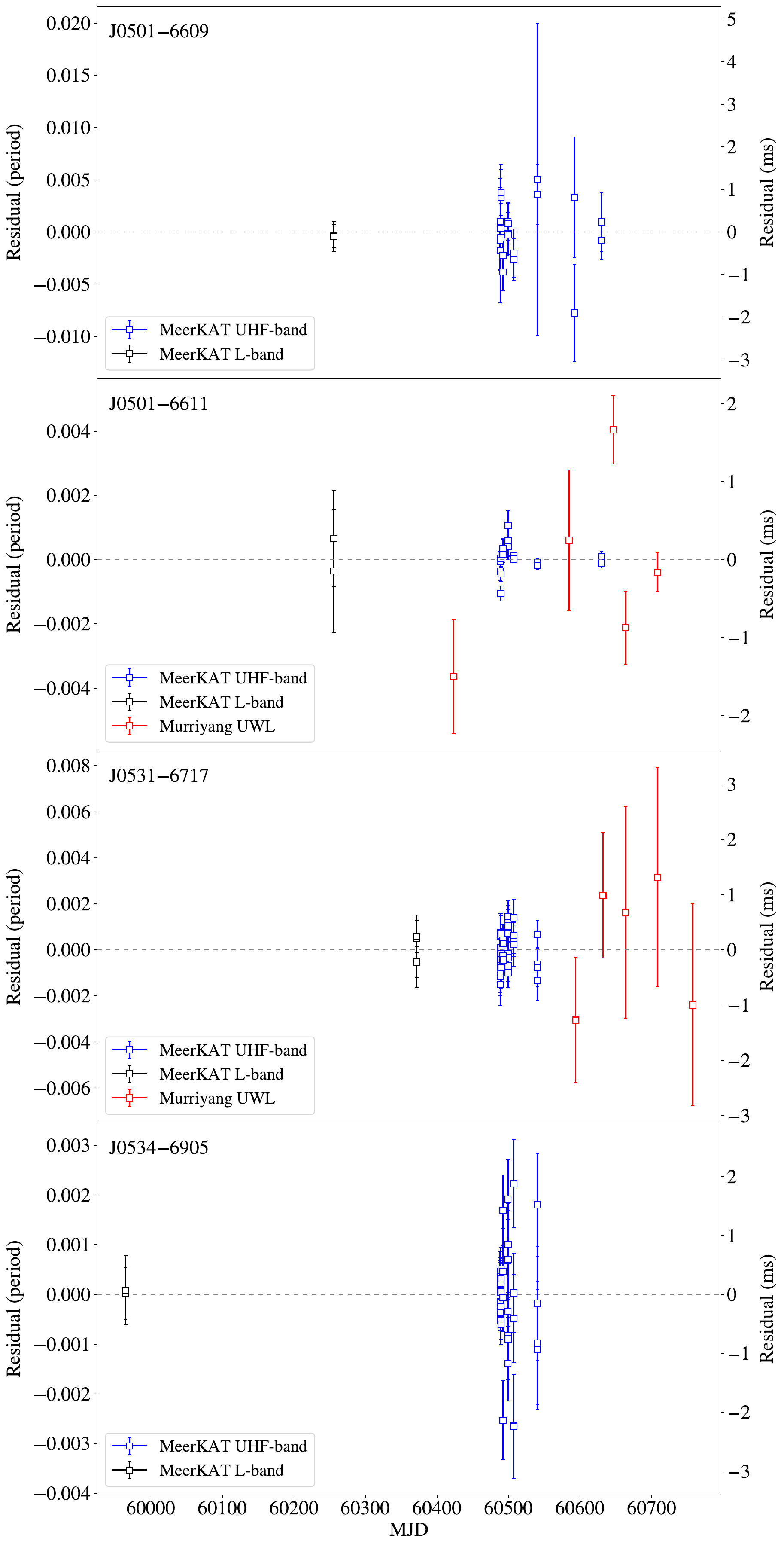}
\caption{The \textsc{tempo2} timing residuals for PSRs\,J0501$-$6609, J0501$-$6611, J0531$-$6717, and J0534$-$6605 (refer to \autoref{table: UHF-band campaign}). These pulsars were part of the second timing campaign.}
\label{figure: timing residuals, second campaign}
\end{figure}

\subsection{Search discoveries} \label{sec : search discoveries}

The 18 new pointings of the TRAPUM LMC Survey yielded 12 new radio pulsars. A summary of the pulsar properties and the integrated discovery pulse profiles are provided in \autoref{table: Discoveries} and \autoref{figure: pulse profiles}, respectively. Their spin periods range from 245 to 2238\,ms. The Milky Way’s DM contribution towards the LMC is around 53\,$pc \, cm^{-3}$ based on the NE2001 model \citep{Cordes2004}, and about 58\,$pc \, cm^{-3}$ using the YMW16 model \citep{Yao2017}. Our new discoveries have DMs ranging from 62 to 326\,$pc \, cm^{-3}$, exceeding the Galactic contribution towards the LMC and thus confirming their association with the LMC. All pulsars targeted in the timing campaigns were successfully re-detected, except for the incoherent beam pulsar (IB pulsar) discovered in Pointing\,1 (see Paper\,I), which was not detected despite partial overlap between the primary beams of Pointing\,16 and Pointing\,1.

The pulse widths were determined by creating noise-free templates using the \texttt{paas} tool from \textsc{psrchive}, and simulating 1000 noise-added profiles based on off-pulse noise statistics. Each profile was fitted with von Mises functions using the \texttt{fitvonMises} routine from \textsc{psrsalsa}\footnote{\href{https://github.com/weltevrede/psrsalsa}{https://github.com/weltevrede/psrsalsa} accessed on 2025 July 25} \citep{Weltevrede2016} to measure the widths at 50 per cent, ${\it W}_{50}$, of the peak intensity. This process also allows an uncertainty to be estimated, and the final width for each pulsar was taken as the average of the simulated measurements. The discovery flux densities of the new pulsars, uncorrected for positional offsets within the CBs, were calculated using the modified radiometer equation \citep{Dewey1985, Lorimer2004}, as described in more detail in Paper\,I:

\begin{equation}
{\it S} = \frac{\betait \: S/N \: ({\it T}_{sys}+{\it T}_{sky})}  {{\it G} \sqrt{{\it n}_{pol}\:{\it t}_{obs}\:\Delta \nu}} \sqrt{\frac{\it W}{{\it P}-{\it W}}}.
\label{eqn: radiometer}
\end{equation}
We used the discovery S/N value (see \autoref{table: Discoveries}), ${\it T}_{sys}=18$\,K \citep{Bailes2020}, ${\it T}_{sky}=4.6$\,K \citep{Price2016, Zheng2017}, and $\betait=1.0$ for 8-bit digitisation \citep{Kouwenhoven2001}. A gain of ${\it G} = 1.84$ K Jy$^{-1}$ was adopted, corresponding to MeerKAT’s core dishes, with ${\it n}_{\mathrm{pol}} = 2$ polarisations. The observation length and total bandwidth are denoted by ${\it t}_{obs}$ and $\Delta \nu$, respectively, and the pulse width \textit{W} was taken as the pulsar's ${\it W}_{50}$. The resulting flux densities were then converted to flux densities at 1400\,MHz, ${\it S}_{\text{1400\,MHz}}$, assuming an average spectral index for the radio pulsar population of $-1.6$ \citep{Jankowski2018}. 

\subsubsection{PSR\,J0452$-$6921}

PSR\,J0452$-$6921 was discovered with a folded S/N of 16.7, from the full 2\,h observation of Pointing\,14. The pulsar was discovered within a beam of the central tiling and also detected in the 2\,h data of two surrounding beams, enabling its localisation using \textsc{seeKAT}. PSR\,J0452$-$6921 has a DM of 325.52\,$pc \, cm^{-3}$, making it the extragalactic radio pulsar with the highest DM discovered to date. The ${\it W}_{50}$ is $20 \pm 1\,ms$, which corresponds to a duty cycle of 2.9 per cent.

\subsubsection{PSR\,J0454$-$6927}

Also discovered within Pointing\,14, PSR\,J0454$-$6927 has a period of 2238\,ms, the longest period among all known extragalactic radio pulsars. It was discovered with a folded S/N of 14.9 within a beam of the central tiling in the complete 2\,h observation. The pulsar was also detected in the complete 2\,h data of two surrounding beams, allowing its localisation using \textsc{seeKAT}. The ${\it W}_{50}$ is $77 \pm 8\,ms$, which corresponds to a duty cycle of 3.5 per cent.

\subsubsection{PSR\,J0458$-$7024}

PSR\,J0458$-$7024 was discovered with a folded S/N of 34.2 within a beam of the central tiling in the complete 2\,h observation of Pointing\,21. It is the brightest detection in our survey so far with a ${\it S}_{\text{1400\,MHz}}$ of 34.2 $\upmu Jy$. The pulsar was further detected in the complete 2\,h data of five surrounding beams, allowing its localisation using \textsc{seeKAT}. The ${\it W}_{50}$ is $24.4 \pm 0.9\,ms$, which corresponds to a duty cycle of 4.6 per cent.

\subsubsection{PSR\,J0501$-$6609}

PSR\,J0501$-$6609 was detected with a folded S/N of 14.7 in the full 2\,h observation of Pointing\,10. It was found within a central tiling beam, with only one additional detection in the 2\,h data of the surrounding beams. As a result, we were unable to achieve a more precise localisation than that provided by the CB in which it was detected. A more accurate localisation was obtained from the second timing campaign where we obtained a phase-connected timing solution for this pulsar (see \autoref{table: UHF-band campaign}). With a period of 245.2\,ms, it is the fastest-spinning pulsar discovered in the TRAPUM LMC Survey up to now. The ${\it W}_{50}$ is $4.9 \pm 0.3\,ms$, which corresponds to a duty cycle of 2 per cent. 

\subsubsection{PSR\,J0501$-$6611}

PSR\,J0501$-$6611, with a folded S/N of 24.2, was also discovered within a central tiling beam of the full 2\,h observation of Pointing\,10. It was further detected in two other surrounding beams from the 2\,h data and we could therefore obtain a better localisation using \textsc{seeKAT}. PSR\,J0501$-$6611 was included in the second timing campaign and has a phase-connected timing solution and an updated position (see \autoref{table: UHF-band campaign}). With a DM of 68.63\,$pc \, cm^{-3}$, this pulsar has the lowest DM among those discovered in the TRAPUM LMC Survey so far. The ${\it W}_{50}$ is $19 \pm 1\,ms$, which corresponds to a duty cycle of 4.6 per cent. 

\subsubsection{PSR\,J0501$-$6744}

PSR\,J0501$-$6744 was discovered with a folded S/N of 13.2, and it was found within a central tiling beam of Pointing\,15 in the full 2\,h observation. Unfortunately, PSR\,J0501$-$6744 was detected in only that beam, preventing a more precise localisation beyond the CB in which it was found. The ${\it W}_{50}$ is $28 \pm 3\,ms$, which corresponds to a duty cycle of 3.1 per cent. 

\begin{table*}
\centering
\caption{Properties of the 12 newly discovered pulsars obtained using our search pipeline (see \autoref{section: Search}), including DMs and spin periods derived using \textsc{pulsarX}, and flux densities and pulse widths measured at 50 per cent of the peak intensity (see \autoref{sec : search discoveries}). For the ${\it W}_{50}$ values, since we are fitting von Mises functions to only a few bins (see \autoref{figure: pulse profiles}), the reported errors are likely underestimated. The quoted uncertainties in parentheses refer to the last significant digit. Positional uncertainties are the 2$\sigma$ errors output by \textsc{seeKAT}, except for pulsars marked with an asterisk (*), where the positional errors correspond to the sizes of the CB in which they were discovered (see \autoref{tab:observation parameters}). This paper presents the timing solutions for PSRs\,J0501$-$6609, J0501$-$6611, and J0531$-$6717, with a more accurate position for PSR\,J0501$-$6609 which established its J2000 designation presented here (see \autoref{table: UHF-band campaign}). We note that some pulsars have ${\it S}_{\text{1400\,MHz}}$ values below the theoretical minimum flux density of 6.3\,$\upmu$Jy reported in Paper\,I, due to having duty cycles smaller than the 2.5 per cent assumed in that calculation.}
\label{table: Discoveries}
\begin{tabular}{lllllcccc} 
\hline
\textbf{Pulsar} & \textbf{RA} & \textbf{Dec} & \textbf{DM} & \textbf{Period} & \textbf{Epoch} & \textbf{S/N} & \boldmath{${\it W}_{50}$} & \boldmath{${\it S}_{\text{1400\,MHz}}$} \\
\text{(J2000)} & \text{(J2000)} & \text{(J2000)} &  \text{($pc \, cm^{-3}$)} & \text{(ms)} & (MJD) &  & \text{(ms)} & \text{($\upmu Jy$)}\\
\hline
\addlinespace[0.1cm]

J0452$-$6921 & 04$^{\rm h}$52$^{\rm m}$53\fs86 $\substack{+4.2 \\ -2.1}$ & $-$69\textdegree{}21\arcmin16\farcs69 $\substack{+6.0 \\ -5.0}$ & 325.52(87) & 690.15853(32) & 60509.1 & 16.7 & $20(1)$ & 9.2\\ \addlinespace[0.2cm]

J0454$-$6927 & 04$^{\rm h}$54$^{\rm m}$29\fs86 $\substack{+1.7 \\ -2.7}$ & $-$69\textdegree{}27\arcmin13\farcs09 $\substack{+5.0 \\ -3.0}$ & 103(6) & 2238.4607(67) & 60509.1 & 14.9 & $77(8)$ & 9.0\\ \addlinespace[0.2cm]

J0458$-$7024 & 04$^{\rm h}$58$^{\rm m}$25\fs34 $\substack{+0.6 \\ -0.6}$ & $-$70\textdegree{}24\arcmin15\farcs30 $\substack{+1.0 \\ -1.0}$ & 118.10(57) & 533.70671(17) & 60750.4 & 34.2 & $24.4(9)$ & 21.8 \\ \addlinespace[0.2cm]

J0501$-$6609* & 05$^{\rm h}$00$^{\rm m}$55\fs53 & $-$66\textdegree{}09\arcmin46\farcs20 & 128.10(26) & 245.200163(35) & 60255.7 & 14.7 & $4.9(3)$ & 6.1\\ \addlinespace[0.2cm]

J0501$-$6611 & 05$^{\rm h}$01$^{\rm m}$18\fs83$\substack{+0.2 \\ -14.5}$ & $-$66\textdegree{}11\arcmin43\farcs29 $\substack{+42.0 \\ -3.0}$ & 68.63(54) & 411.82478(12) & 60255.7 & 24.2 & $19(1)$ & 15.5\\ \addlinespace[0.2cm]

J0501$-$6744* & 05$^{\rm h}$01$^{\rm m}$24\fs81   & $-$67\textdegree{}44\arcmin23\farcs70 & 96(2)  & 887.12646(85) & 60608.3 & 13.2 & $28(3)$ & 6.3\\ \addlinespace[0.2cm]

J0501$-$6750 & 05$^{\rm h}$01$^{\rm m}$27\fs38 $\substack{+1.1 \\ -1.4}$ & $-$67\textdegree{}50\arcmin00\farcs10 $\substack{+3.0 \\ -3.0}$ & 86.80(29) & 263.845224(42) & 60608.3 & 14.2 & $4.1(4)$ & 4.8\\ \addlinespace[0.2cm]

J0505$-$6530 & 05$^{\rm h}$05$^{\rm m}$15\fs78 $\substack{+0.6 \\ -0.8}$ & $-$65\textdegree{}30\arcmin43\farcs99 $\substack{+3.0 \\ -4.0}$ & 85(2) & 915.88844(85) & 60750.5 & 16.8 & $30(2)$ & 9.0\\ \addlinespace[0.2cm]

J0525$-$6950 & 05$^{\rm h}$25$^{\rm m}$41\fs31 $\substack{+0.0 \\ -0.6}$ & $-$69\textdegree{}50\arcmin57\farcs50 $\substack{+3.0 \\ -1.0}$ & 245.35(40) & 719.44816(15) & 60608.4 & 28.6 & $15.8(8)$ & 11.4
\\ \addlinespace[0.1cm]

J0527$-$6935 & 05$^{\rm h}$27$^{\rm m}$19\fs92 $\substack{+1.0 \\ -0.0}$ & $-$69\textdegree{}35\arcmin05\farcs70 $\substack{+2.0 \\ -3.0}$ & 101.69(32) & 601.42403(10) & 60608.4 & 30.0 & $10.0(4)$ & 10.4
\\ \addlinespace[0.1cm]

J0531$-$6717 & 05$^{\rm h}$31$^{\rm m}$04\fs23 $\substack{+1.0 \\ -0.9}$ & $-$67\textdegree{}17\arcmin35\farcs60 $\substack{+3.0 \\ -2.0}$ & 79.50(25) & 416.746779(57) & 60371.5 & 26.1 & $8.9(3)$ & 11.3

\\ \addlinespace[0.1cm]

J0537$-$6957* & 05$^{\rm h}$37$^{\rm m}$59\fs04 & $-$69\textdegree{}57\arcmin28\farcs00 & 61.90(45) & 380.96915(10) & 60709.5 & 21.7 & $8.9(4)$ & 10.8

\\ \addlinespace[0.1cm]
\hline
\end{tabular}
\end{table*}

\subsubsection{PSR\,J0501$-$6750}

PSR\,J0501$-$6750, with a folded S/N of 14.2, was discovered in the same pointing as PSR\,J0501$-$6744. The pulsar was found within a beam of the central tiling of Pointing\,15, and further detected in the complete 2\,h data of three surrounding beams, enabling its localisation using \textsc{seeKAT}. The ${\it W}_{50}$ is $4.1 \pm 0.4\,ms$, which corresponds to a duty cycle of 1.6 per cent. 

    \subsubsection{PSR\,J0505$-$6530}

PSR\,J0505$-$6530 was discovered with a folded S/N of 16.8 within a beam of the central tiling in the complete 2\,h observation of Pointing\,22. It was further detected in the complete 2\,h data of five surrounding beams, allowing its localisation using \textsc{seeKAT}. The ${\it W}_{50}$ is $30 \pm 2\,ms$, which corresponds to a duty cycle of 3.2 per cent.

\subsubsection{PSR\,J0525$-$6950}

PSR\,J0525$-$6950 was discovered with a folded S/N of 28.6 in the complete 2\,h observation of Pointing\,16. It was found within a beam of the central tiling, and further detected in the complete 2\,h data of three surrounding beams, allowing its localisation with \textsc{seeKAT}. The ${\it W}_{50}$ is $15.8 \pm 0.8\,ms$, which corresponds to a duty cycle of 2.2 per cent.

\subsubsection{PSR\,J0527$-$6935}
PSR\,J0527$-$6935, with a folded S/N of 30.0, the highest among the new discoveries, was also detected within a central tiling beam during the full 2\,hour observation of Pointing\,16. It was further detected in the complete 2\,h data of six surrounding beams, possibly in part due to its brightness, enabling its localisation using \textsc{seeKAT}. The ${\it W}_{50}$ is $10.0 \pm 0.4\,ms$, which corresponds to a duty cycle of 1.7 per cent. 

\subsubsection{PSR\,J0531$-$6717}

PSR\,J0531$-$6717 was discovered with a folded S/N of 26.1 in the complete 2\,h observation of Pointing\,12. It was found within a beam of the central tiling and detected in the full 2\,h observations of three surrounding beams, enabling its localisation using \textsc{seeKAT}. PSR\,J0531$-$6717 was included in the second timing campaign, for which we now have a phase-connected timing solution and an updated position (see \autoref{table: UHF-band campaign}). The ${\it W}_{50}$ is $8.9 \pm 0.3\,ms$, which corresponds to a duty cycle of 2.1 per cent.

\subsubsection{PSR\,J0537$-$6957}

PSR\,J0537$-$6957 was discovered with a folded S/N of 26.1 in the complete 2\,h observation of Pointing\,19. The pulsar was found within a beam of the central tiling in the 2\,h data. It was detected in only one additional beam, so we could not obtain a better localisation using \textsc{seeKAT} than the CB in which it was found. The ${\it W}_{50}$ is $8.9 \pm 0.4\,ms$, which corresponds to a duty cycle of 2.1 per cent.

\subsection{Improved localisation of PSR\,J0536$-$6923} \label{section: Localisation of PSR J0537-69} 

The discovery of PSR\,J0536$-$6923 (previously PSR\,J0537$-$69) was reported in \cite{Ridley2013} from a 2\,h observation of the High Resolution LMC Survey using Murriyang and the Berkeley–Parkes–
Swinburne data recorder (BPSR) receiver with a S/N of 8.5. In Pointing\,3, as reported in Paper\,I, we detected PSR\,J0536$-$6923 with a S/N of 79, and upon further inspection, we identified it within four surrounding CBs of the central tiling. Using \textsc{seeKAT}, we obtained a better localisation, placing it at a right ascension and declination of 05$^{\rm h}$36$^{\rm m}$47\fs32 $\substack{+0.4 \\ -0.0}$ and $-$69\textdegree{}23\arcmin45\farcs50 $\substack{+0.0 \\ -1.0}$. The integrated pulse profile of PSR\,J0536$-$6923 is shown in \autoref{figure: 0537 pulse profile} and the ${\it W}_{50}$ is $2.43 \pm 0.05\,ms$, which corresponds to a duty cycle of 2.2 per cent.

\subsection{Timing solutions} \label{section: timing solutions}

\begin{table*}
\centering
\caption{Timing solutions from the first campaign, based on MeerKAT L-band and Murriyang UWL observations, obtained using \textsc{tempo2} by fitting to the observed ToAs weighted by their uncertainties (see \autoref{figure: timing residuals, first campaign}). The derived quantities are listed with uncertainties in parentheses, indicating the error on the final digit. The fixed parameters were held constant during the fit and their uncertainties were not included in the solution. The timing solution for PSR\,J0509$-$6838 is tentative (see \autoref{section: timing solutions}).} 
\label{table: L-band campaign}
\begin{tabular}{lllll}

\hline\hline
\multicolumn{5}{c}{\textbf{Data and Modelling}} \\
\hline
Pulsar name\dotfill & \textbf{J0509$-$6838} & \textbf{J0509$-$6845} & \textbf{J0518$-$6939} & \textbf{J0519$-$6931} \\ 
MJD range\dotfill & 59830.4---60180.3 & 59919.8---60195.6 & 59919.8---60180.3 &  59919.8---60180.3 \\ 
Data span (yr)\dotfill & 0.96 & 0.76 & 0.71 &  0.71 \\ 
Number of ToAs\dotfill & 10 & 38 & 40 & 10 \\
Rms timing residual ($\upmu s$)\dotfill & 1582.6 & 78.4 & 335.9 & 478.6 \\
Reduced $\chi^2$ value \dotfill & 7.7 & 0.9 &  1.1 &  0.8 \\
\hline
\multicolumn{5}{c}{\textbf{Measured quantities}} \\ 
\hline
Right ascension (J2000, hh:mm:ss)\dotfill &  05:09:55.10(42) & 05:09:51.93(4) & 05:18:16.82(9) &  05:19:09.32(40) \\ 
Declination (J2000, deg:arcmin:arcsec)\dotfill & $-$68:39:18.20(15) & $-$68:45:19.29(13) & $-$69:39:38.85(22) &  $-$69:31:27.25(85) \\ 
Pulse frequency, $\nu$ (s$^{-1}$)\dotfill & 3.5874009867(14) & 
3.25524856937(57) & 3.02838957031(86) & 2.6473627960(34) \\ 
First derivative of pulse frequency, $\dot{\nu}$ (s$^{-2}$)\dotfill & $-2.35437(5)\times 10^{-12}$ & $-$6.81(6)$\times 10^{-15}$  &  $-$1.1849(8)$\times 10^{-13}$ & $-$1.750(3)$\times 10^{-13}$ \\
\hline
\multicolumn{5}{c}{\textbf{Fixed quantities}} \\ 
\hline
Dispersion measure, DM (cm$^{-3}$pc)\dotfill & 
147.5(10) & 
90.9(11) & 254.2(12) & 82.7(13) \\ 
Epoch of pulse frequency (PEPOCH)\dotfill & 59968.5 & 59919.9 & 59919.8 & 59923.6 \\
\hline
\multicolumn{5}{c}{\textbf{Derived quantities}} \\
\hline
Characteristic age (kyr) \dotfill & 24 &  7574 & 405 & 240 \\
$\log_{10}$(Surface magnetic field strength, G) \dotfill & 12.9 & 11.7 & 12.3 & 12.5 \\
$\log_{10}$(Spin-down luminosity, $\dot{E}$, erg\,s$^{-1}$)  & 35.5 & 32.9 & 34.2 &  34.3 \\
\hline
\hline
\end{tabular}
\end{table*}

\begin{table*}
\centering
\caption{Timing solutions from the second campaign, based on MeerKAT UHF-band and Murriyang UWL timing observations, together with the L-band discovery data, were obtained using \textsc{tempo2} by fitting to the observed ToAs weighted by their uncertainties (see \autoref{figure: timing residuals, second campaign}). The remaining details are as described in the caption of \autoref{table: L-band campaign}.} 
\label{table: UHF-band campaign}
\begin{tabular}{lllll}

\hline\hline
\multicolumn{5}{c}{\textbf{Data and Modelling}} \\
\hline
Pulsar name\dotfill & \textbf{J0501$-$6609} & \textbf{J0501$-$6611} & \textbf{J0531$-$6717} & \textbf{J0534$-$6905} \\ 
MJD range\dotfill & 60371.5---60757.3 & 60255.7---60708.2 & 60488.3---60708.3 &  59964.5---60540.3 \\ 
Data span (yr)\dotfill & 1.02 & 1.24 & 1.06 & 1.58 \\ 
Number of ToAs\dotfill & 24 & 25 & 45 & 38 \\
Rms timing residual ($\upmu s$)\dotfill & 409.5 & 140.9 & 321.4 &  573.9 \\
Reduced $\chi^2$ value \dotfill & 0.7 & 2.7 & 0.9 & 1.6 \\
\hline
\multicolumn{5}{c}{\textbf{Measured quantities}} \\ 
\hline
Right ascension (J2000, hh:mm:ss)\dotfill &  05:01:04.73(2) & 05:01:27.55(1) & 05:31:03.44(32) &  05:34:45(8)\\ 
Declination (J2000, deg:arcmin:arcsec)\dotfill & $-$66:09:14.32(28) & $-$66:11:22.27(15) & $-$67:17:30.86(60) & -69:05:02(45) \\ 
Pulse frequency, $\nu$ (s$^{-1}$)\dotfill & 4.0783283518(13) & 
2.428233841899(70) & 2.39953798713(37) & 
1.186543165(31)
 \\ 
First derivative of pulse frequency, $\dot{\nu}$ (s$^{-2}$)\dotfill & $-7.3(3)\times 10^{-15}$ & $-$1.632(1)$\times 10^{-14}$  &  $-$6.11(5)$\times 10^{-15}$ & $-$2.9(1)$\times 10^{-14}$ \\
\hline
\multicolumn{5}{c}{\textbf{Fixed quantities}} \\ 
\hline
Dispersion measure, DM (cm$^{-3}$pc)\dotfill & 
128.9(9) & 
69.0(14) & 78.7(15) & 243(3) \\ 
Epoch (MJD)\dotfill & 60499.3 & 60489.4 & 60492.3 & 60488.4 \\
Right ascension (hh:mm:ss)\dotfill & -- & -- & -- & -- \\ 
Declination (dd:mm:ss)\dotfill & -- & -- & -- & -- \\
\hline
\multicolumn{5}{c}{\textbf{Derived quantities}} \\
\hline
Characteristic age (kyr) \dotfill & 8848 &  2360 & 6225 & 642 \\
$\log_{10}$(Surface magnetic field strength, G) \dotfill & 11.5 & 12.0 & 11.8 & 12.6 \\
$\log_{10}$(Spin-down luminosity, $\dot{E}$, erg\,s$^{-1}$)  & 33.1 & 33.2 & 32.8 &  33.1 \\
\hline
\hline
\end{tabular}
\end{table*}

\begin{figure*}
\includegraphics[width=1.75\columnwidth]{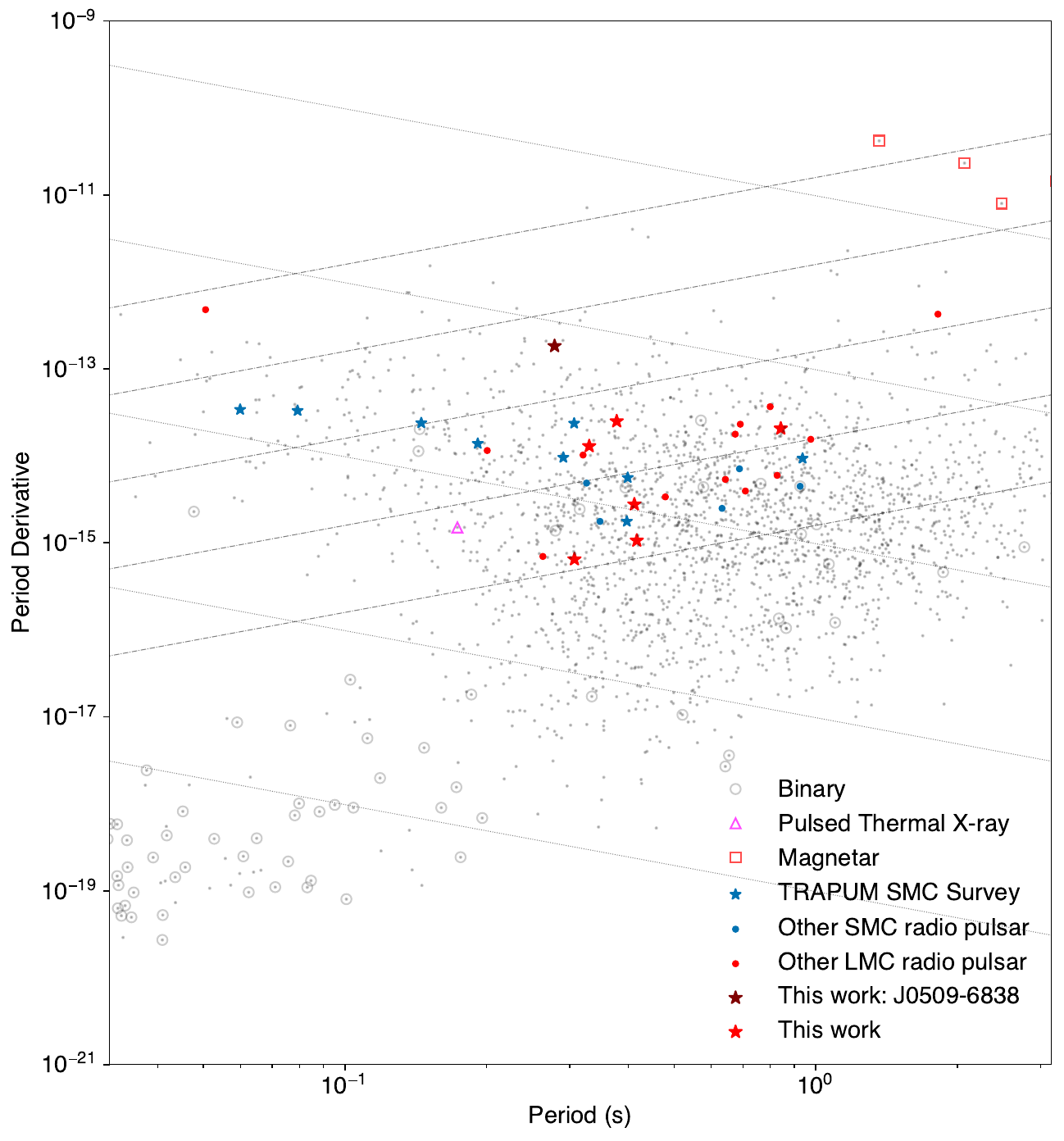}
\caption{Pulsars plotted by spin period and period derivative, forming the characteristic $P{-}\dot{P}$ diagram. Galactic pulsars are shown as filled grey circles, with those in binary systems marked by surrounding open circles. The star markers in blue represent the results from the TRAPUM SMC Survey \protect\citep{Carli2024b}, while the ones in red represent this work. PSR\,J0509$-$6838 is included based on a provisional solution. The filled blue and red circles are the already published SMC and LMC radio pulsars, respectively. The diagonal lines represent constant values of characteristic age and surface magnetic field. This plot was made using the Python package \textsc{PSRQpy} \citep{psrqpy} with data from the  ATNF Pulsar Database, PSRCat v2.6.5 \protect\citep{Manchester2005}.}
\label{figure: p-pdot}
\end{figure*}

The timing parameters obtained from the first and second timing campaigns are shown in \autoref{table: L-band campaign} and \autoref{table: UHF-band campaign}, respectively. We have phase-connected timing solutions for seven out of the eight timed pulsars, all of which indicate that they are isolated. PSR\,J0509$-$6838 could not be phase-connected using the ten ToAs spanning approximately 350 days, including the discovery observation, and its timing solution therefore remains preliminary. Its reduced chi-squared is noticeably higher than for the other sources, and this seems to stem from small systematic trends in the residuals that we cannot yet model reliably with the current data set. Discovery observations were also included for several pulsars to extend the timing baseline and improve their timing. The timing residuals generated from the \textsc{tempo2} best fits are shown in \autoref{figure: timing residuals, first campaign} for the first campaign and \autoref{figure: timing residuals, second campaign} for the second campaign. The periods and period derivatives derived from the timing campaigns are shown in the $P{-}\dot{P}$ diagram (\autoref{figure: p-pdot}), alongside the known pulsar population. 

\begin{figure*}
\includegraphics[width=2\columnwidth]{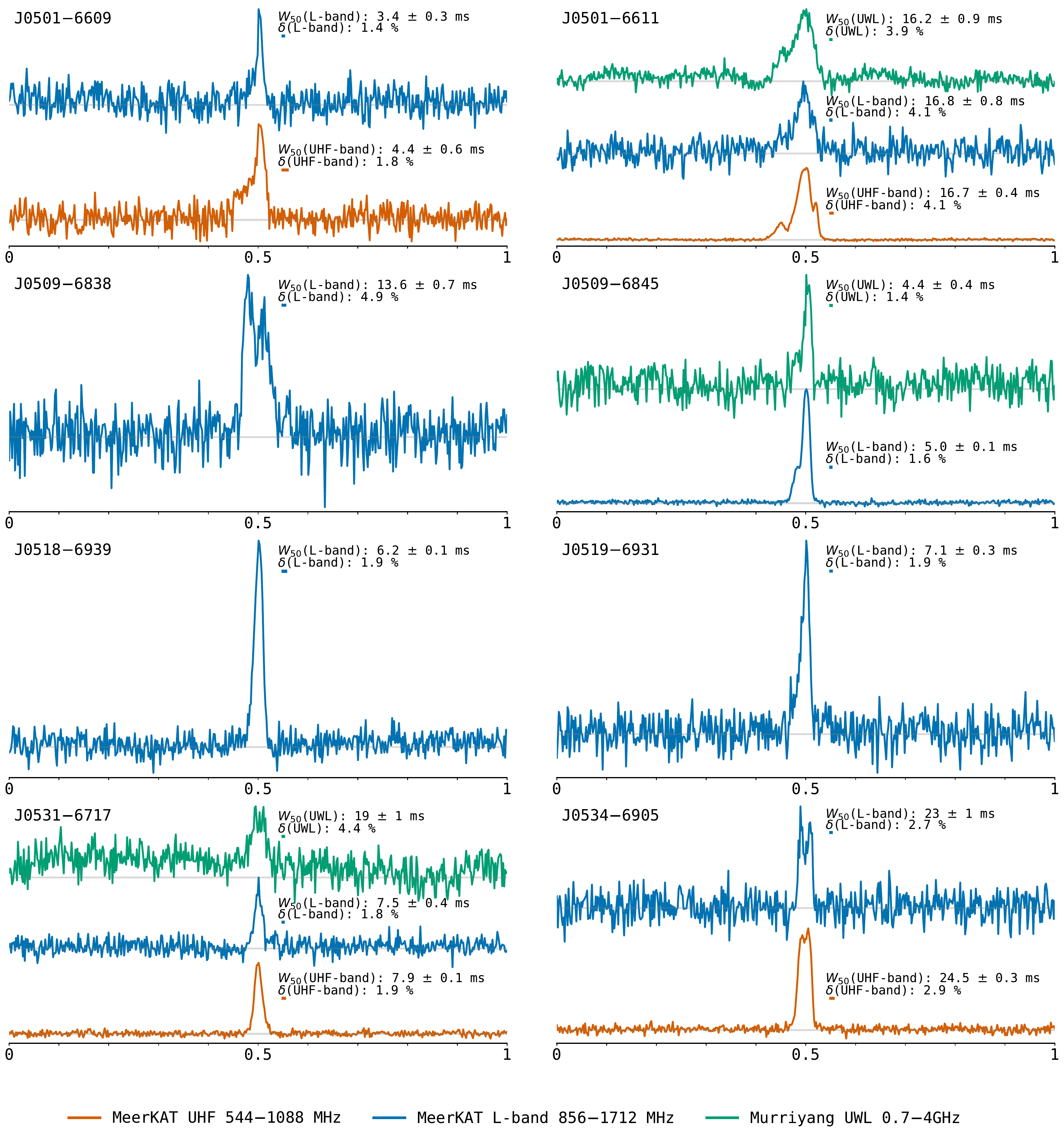}
\caption{The summed pulse profiles from the timing campaign, divided into 512 phase bins and arbitrarily centred at phase 0.5 for ease of comparison, are shown alongside their respective 50 per cent widths and duty cycles. Rotational phase spans 0 to 1, and all profiles have been normalised to a unit peak, with the zero baseline marked by the grey horizontal line. For the pulsars that were not timed at MeerKAT’s L-band, namely PSRs\,J0501$-$6609, J0501$-$6611, J0531$-$6717 and J0534$-$6905, the L-band pulse profiles shown are taken from their respective subbanded discovery beam data, each also consisting of 512 phase bins, in contrast to the 128 phase bins shown in \autoref{figure: pulse profiles}. The width of the horizontal bar represents the effective time resolution of the system relative to each pulsar’s spin period.}
\label{figure: timing pulse profiles}
\end{figure*}

The pulse profiles of the timed pulsars, summed over multiple epochs, are presented in \autoref{figure: timing pulse profiles}. For some pulsars, multiple profiles are shown corresponding to observations at different frequencies. PSR\,J0501$-$6609 exhibits a broader pulse profile in the UHF-band, where it appears as a single component with a leading shoulder, which evolves into a narrower and more defined single component at L-band. PSR\,J0501$-$6611 shows a three-component pulse profile in the UHF-band, consisting of a core and two conal components \citep{Rankin1983}, which evolve into a narrower two-component pulse shape at higher frequencies in the L- and UWL-bands. PSR\,J0509$-$6838 exhibits a two-component profile in the L-band, characterised by a steep leading edge and a more gradual trailing edge, with its relatively large width and duty cycle indicating emission from a broad region. The pulse profile of PSR\,J0509$-$6845 exhibits a narrow main component with a leading shoulder, with the pulse shape remaining stable from the L-band to the UWL-band while the overall pulse width becomes slightly narrower. PSRs\,J0518$-$6939 and J0519$-$6931 both exhibit narrow, single-component pulse profiles at L-band, with J0519$-$6931 showing a slight leading shoulder. PSR\,J0531$-$6717 exhibits a single-component pulse profile in the UHF-band that evolves into a wider pulse shape at higher frequencies in the UWL-band. PSR\,J0534$-$6905 exhibits a single-component profile with a hint of a double peak at UHF-band, which narrows at L-band to reveal a clear double-peaked structure.

\begin{center}
\begin{figure*}
\begin{tabular}{c}
\includegraphics[width=2\columnwidth]{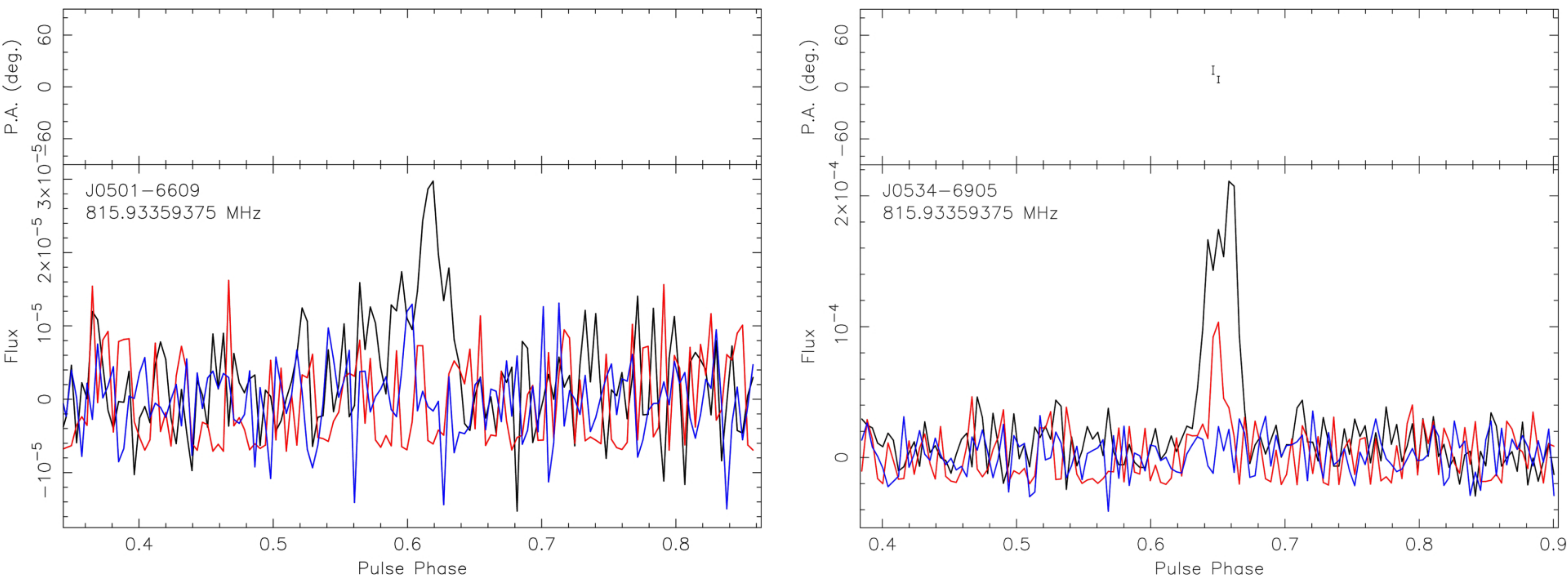} \\
\end{tabular}
\caption{The polarisation profiles, using 256 phase bins, made using \textsc{psrchive}’s \texttt{psrplot}, are shown for PSRs\,J0501$-$6609 and J0534$-$6905, with only the latter corrected for Faraday rotation using its RM (see \autoref{section: Polarisation profiles}). Rotational phase spans 0 to 1, and the y-axis shows flux density in arbitrary units. The black, blue, and red lines represent the total intensity, circular polarisation, and linear polarisation, respectively.}
\label{figure: polarisation profiles}
\end{figure*}
\end{center}

\subsection{Polarisation profiles} \label{section: Polarisation profiles}

During the second timing campaign, at UHF-band, some timing observations of PSRs\,J0501$-$6609 and J0534$-$6905 were recorded using the Pulsars and Transients User Supplied Equipment (PTUSE; \citealt{Bailes2020}) alongside APSUSE. MeerKAT array calibration allows PTUSE to deliver calibrated Stokes information across frequency (as described in \citealt{Serylak2021}). From these we construct the polarisation profiles of these two pulsars after applying a final frontend correction using \texttt{pac} with the \texttt{-XP} option within the \textsc{psrchive} tool. The simultaneous PTUSE data were recorded in fold mode, generating \textsc{psrchive} archive files, each containing multiple 8\,s subintegrations of folded data.  The individual files were combined into a single archive using \textsc{psrchive}'s \texttt{psradd} and subsequently cleaned of RFI. The \texttt{rmsynth} tool from \textsc{psrsalsa} was used to perform a rotation measure (RM) search over a range of $\pm 300\,\mathrm{rad\,m^{-2}}$, consistent with the RM values of some known LMC pulsars, which range between $-246$ and $+85\,\mathrm{rad\,m^{-2}}$ \citep{Johnston2022}.

For PSR\,J0534$-$6905, we obtained a RM of $-20.2\pm6.8$\,rad\,$m^{-2}$. We were unable to constrain the RM for PSR\,J0501$-$6609, most probably due to a low S/N pulse detection and a low presence of linear polarisation.  \autoref{figure: polarisation profiles} shows the polarisation profiles of PSRs\,J0501$-$6609 and J0534$-$6905.

\section{Discussion and conclusions} \label{section: discussion and conclusions}

The discovery of twelve new radio pulsars increases the total number of known radio pulsars in the LMC to 44, marking an 80 per cent increase since the start of the TRAPUM LMC Survey. We have used the ASKAP 888\,MHz radio continuum survey image of the LMC \citep{Pennock2021} to cross-check the coordinates of the pulsars, and they do not appear to be associated with any of the sources listed in \autoref{appendix: Targeted sources} of the Appendix.

PSR\,J0537$-$6957 has a DM of 61.90\,$\mathrm{pc\,cm^{-3}}$, making it the extragalactic pulsar with the second lowest DM after PSR\,J0451$-$67, also in the LMC, which has a DM of 45\,$\mathrm{pc\,cm^{-3}}$. This may suggest that the source lies within a low-electron-density region or towards the near side of the LMC. By contrast, with a DM of 325.25\,$pc \, cm^{-3}$, PSR\,J0452$-$6921 has the highest known DM among extragalactic radio pulsars discovered to date. The high DM value may be attributed to the pulsar’s location in a region of bright radio emission, where dust and gas are likely abundant (see Pointing\,14 in \autoref{appendix: beam maps}). Within the same pointing, PSR\,J0454$-$6927 was found to have the longest period of any extragalactic radio pulsar identified so far, 2.24\,s. The second longest period is that of PSR\,J0534$-$6845, also within the LMC, which has a period of 1.82\,s.

PSR\,J0527$-$6935 has a DM of 101.7\,$pc \, cm^{-3}$ and a period of 601.4\,ms, which is almost exactly three times that of the known pulsar PSR\,J0535$-$6935, which has a period of 200.5\,ms and DM of 93.7\,$pc \, cm^{-3}$ \citep{Crawford2001}. The discovery beam of PSR\,J0527$-$6935 lies approximately 40\,arcminutes from the nominal position of PSR\,J0535$-$6935, which has a positional uncertainty radius of 7\,arcminutes, the same as the radius of the discovery beam using Murriyang. Even without a clear overlap between the sizes and positions of the discovery beams of the two pulsars, we examined Murriyang’s archives for observations of PSR\,J0535$-$6935. This pulsar was detected approximately 25 times in about 50 observations between 1998 and 2004 \citep{Manchester2006}, and the search was intended to check whether it is in fact the third harmonic of PSR\,J0527$-$6935. However, we did not detect PSR\,J0527$-$6935 in the data and thus concluded that they are indeed two distinct pulsars.

The detected known pulsars include PSRs\,J0455$-$6951, J0456$-$7031, J0529$-$6652, J0534$-$6703, and J0543$-$6851, all previously well localised, as well as J0535$-$66, which was detected only within the incoherent beam of Pointing\,12. PSR\,J0535$-$66 is also not associated with the HMXB\,1A0535$-$668 (also known as A0538$-$66, \citealt{White1978}), where X-ray pulsations have been rediscovered recently \citep{Ducci2025}. We did not detect any pulsed radio emission from 1A0535$-$668. PSR\,J0529$-$6652 shows nulling in its emission \citep{Crawford2013, Johnston2022}. Nulling refers to the sudden disappearance of pulsed radio emission for one or more pulse periods, after which the pulsar resumes emitting as normal (e.g. \citealt{Backer1970, Ng2020}). In our 2\,hour search observation, episodic nulling was also observed, with the longest nulls each lasting approximately 12 minutes. The known pulsars that were observed (see \autoref{appendix: Targeted sources} of the Appendix) but not detected in our search observations have poorly constrained positions and they are most likely located outside of our CB tiling area.

From the timing of PSR\,J0509$-$6838, discovered in Pointing\,2 (see Paper\,I), the preliminary derived parameters suggest a young pulsar, with a characteristic age of $\sim$24 kyr. We cross-checked the MeerKAT 1.3\,GHz radio continuum images of the LMC (Bill Cotton, private communication about unpublished data) for any visible signs of a SNR or a PWN within the error box defined by the best timing localisation using \textsc{seeKAT}, but found none. We also examined eRosita data (Werner Becker, private communication), as young pulsars tend to emit in X-rays \citep{Becker1997, Owen2011}, but without success. From the timing analysis, there are indications that the pulsar may be in a binary system, as phase connection could not be maintained and the spin period was observed to increase, with no evidence of a glitch. Additional timing observations are planned to confirm this possibility.

 The parameters derived from the timing campaigns place the seven timed pulsars among the population of normal, or “slow”, pulsars in the $P{-}\dot{P}$ diagram, alongside most of the previously timed radio pulsars in the Magellanic Clouds (see \autoref{figure: p-pdot}). Notably, PSR\,J0501$-$6609 currently has the largest characteristic age among extragalactic radio pulsars, at 8.85\,Myr. A detailed comparison of pulsars in the LMC and the Milky Way will be undertaken using our full sample.

\section{Acknowledgements} \label{section: Acknowledgements}

We are grateful to the referee for comments that helped to improve this paper.

The MeerKAT telescope is operated by the South African Radio
Astronomy Observatory, which is a facility of the National Research Foundation, an agency of the Department of Science and Innovation. SARAO acknowledges the ongoing advice and calibration of GPS systems by the National Metrology Institute of South Africa (NMISA) and the time space reference systems department of the
Paris Observatory.

TRAPUM observations used the FBFUSE and APSUSE computing clusters for beamforming, data acquisition, storage, and analysis. These clusters were designed, funded, and installed by the Max-Planck-Institut für Radioastronomie and the Max-Planck-Gesellschaft. PTUSE was developed with support from the Australian SKA Office and Swinburne University of Technology.

The Parkes `Murriyang' radio telescope is part of the Australia Telescope National Facility which is funded by the Australian Government for operation as a National Facility managed by CSIRO (the Commonwealth Scientific and Industrial Research Organisation). We acknowledge the Wiradjuri people as the Traditional Owners of the Observatory site.

This research has made use of the SIMBAD database, operated at CDS, Strasbourg, France \citep{Wenger2000}, and NASA’s \href{https://ui.adsabs.harvard.edu/}{Astrophysics Data System} Bibliographic Services.

This research would not have been possible without the financial support for VP, which came from the National Organising Committee for the International Astronomical Union General Assembly 2024, through the Africa 2024 Scholarship, the African Astronomical Society (AfAS) Seed Research Grant 2025, and the UCT-SKA Doctoral Scholarship. 

\section{Data availability}

The discovery PSRCHIVE archive files of the new pulsars can be found on \textsc{ZENODO} at \href{https://zenodo.org/doi/10.5281/zenodo.17448374}{DOI 10.5281/zenodo.17448374}. The TRAPUM collaboration will share the candidates list upon appropriate request. 

\bibliographystyle{mnras}
\bibliography{references} 

\clearpage

\appendix
\onecolumn
\section{Observations}\label{Appendix A: observations}
The table below summarises the observational setup for pointings 5--22 (refer to \autoref{section: observations}). For each pointing, we list the date of observation, the pointing centre, the number of dishes contributing to both the IBs and CBs, the size of the CBs, and the total duration of each observation. 
\twocolumn
\begin{table*}
\centering
\caption{List of parameters for the 18 MeerKAT pointings. The major and minor axes of the CBs are provided in arcseconds, along with the position angles, which are measured from East through North on the sky \citep{Chen2021, Bezuidenhout2023}.}
\label{tab:observation parameters}
\begin{tabular}{ccccccc}
\hline
\textbf{Observation Date} & \multicolumn{1}{l}{\textbf{Pointing centre}} & \textbf{CB dishes (number of CBs)} & \textbf{IB dishes} & \textbf{CB size} & \textbf{Observation duration}\\ \hline

\begin{tabular}[c]{@{}c@{}}Pointing\,5\\ 2023 Jun 15\end{tabular} & \begin{tabular}[c]{@{}c@{}}05$^{\rm h}$43$^{\rm m}$44\fs69\\ $-$69\textdegree{}06\arcmin41\farcs40\end{tabular} & 44 (285) & 61 & 59.64\arcsec, 22.35\arcsec, $-$22.60$^{\circ}$ & 7179.39\,s\\ \hline

\begin{tabular}[c]{@{}c@{}}Pointing\,6\\ 2023 Jun 15 \end{tabular} & \begin{tabular}[c]{@{}c@{}}05$^{\rm h}$29$^{\rm m}$27\fs89\\ $-$66\textdegree{}05\arcmin36\farcs10\end{tabular} & 44 (286)& 61 & 40.40\arcsec, 22.51\arcsec, 6.23$^{\circ}$ & 7189.43\,s\\ \hline

\begin{tabular}[c]{@{}c@{}}Pointing\,7\\ 2023 Aug 05 \end{tabular} & \begin{tabular}[c]{@{}c@{}}05$^{\rm h}$32$^{\rm m}$24\fs80\\ $-$67\textdegree{}11\arcmin08\farcs70\end{tabular} & 48 (766) & 64 & 70.60\arcsec, 20.82\arcsec, $-$29.90$^{\circ}$ & 7177.51\,s\\ \hline

\begin{tabular}[c]{@{}c@{}}Pointing\,8\\ 2023 Aug 06 \end{tabular} & \begin{tabular}[c]{@{}c@{}}05$^{\rm h}$46$^{\rm m}$33\fs25\\ $-$69\textdegree{}41\arcmin58\farcs70\end{tabular} & 48 (767) & 64 & 25.04\arcsec, 12.93\arcsec, 5.98$^{\circ}$ & 7181.90\,s\\ \hline

\begin{tabular}[c]{@{}c@{}}Pointing\,9\\ 2023 Nov 07 \end{tabular} & \begin{tabular}[c]{@{}c@{}}05$^{\rm h}$32$^{\rm m}$36\fs03\\ $-$70\textdegree{}55\arcmin48\farcs70\end{tabular} & 44 (764) & 56 & 50.92\arcsec, 25.84\arcsec, 0.15$^{\circ}$ & 7156.19\,s\\ \hline

\begin{tabular}[c]{@{}c@{}}Pointing\,10\\ 2023 Nov 07 \end{tabular} & \begin{tabular}[c]{@{}c@{}}04$^{\rm h}$59$^{\rm m}$37\fs91\\ $-$66\textdegree{}21\arcmin28\farcs00\end{tabular} & 44 (766) & 56 & 70.69\arcsec, 25.93\arcsec, $-$23.49$^{\circ}$ & 7146.78\,s\\ \hline

\begin{tabular}[c]{@{}c@{}}Pointing\,11\\ 2024 Mar 02 \end{tabular} & \begin{tabular}[c]{@{}c@{}}05$^{\rm h}$11$^{\rm m}$13\fs58\\ $-$67\textdegree{}21\arcmin59\farcs90\end{tabular} & 44 (766) & 63 & 71.58\arcsec, 24.13\arcsec, $-$26.62$^{\circ}$ & 7154.31\,s\\ \hline

\begin{tabular}[c]{@{}c@{}}Pointing\,12\\ 2024 Mar 02 \end{tabular} & \begin{tabular}[c]{@{}c@{}}05$^{\rm h}$32$^{\rm m}$29\fs83\\ $-$67\textdegree{}16\arcmin38\farcs70\end{tabular} & 44 (767) & 63 & 47.62\arcsec, 24.35\arcsec, 0.57$^{\circ}$ & 7173.12\,s\\ \hline

\begin{tabular}[c]{@{}c@{}}Pointing\,13\\ 2024 Jul 17 \end{tabular} & \begin{tabular}[c]{@{}c@{}}04$^{\rm h}$53$^{\rm m}$03\fs44\\ $-$68\textdegree{}26\arcmin47\farcs60\end{tabular} & 40 (761) & 58 & 64.83\arcsec, 23.46\arcsec, $-$26.57$^{\circ}$ & 	7154.31\,s\\ \hline

\begin{tabular}[c]{@{}c@{}}Pointing\,14\\ 2024 Jul 18  \end{tabular} & \begin{tabular}[c]{@{}c@{}}04$^{\rm h}$52$^{\rm m}$23\fs14\\ $-$69\textdegree{}30\arcmin37\farcs10\end{tabular} & 40 (756) & 58 & 44.38\arcsec, 23.57\arcsec, 0.71$^{\circ}$ & 	7174.38\,s\\ \hline

\begin{tabular}[c]{@{}c@{}}Pointing\,15\\ 2024 Oct 25 \end{tabular} & \begin{tabular}[c]{@{}c@{}}05$^{\rm h}$02$^{\rm m}$00\fs00\\ $-$67\textdegree{}44\arcmin58\farcs00\end{tabular} & 48 (758) & 64 & 42.34\arcsec, 22.48\arcsec, $-$1.15$^{\circ}$ & 7156.19\,s\\ \hline

\begin{tabular}[c]{@{}c@{}}Pointing\,16\\ 2024 Oct 25 \end{tabular} & \begin{tabular}[c]{@{}c@{}}05$^{\rm h}$26$^{\rm m}$39\fs65\\ $-$69\textdegree{}39\arcmin50\farcs40\end{tabular} & 48 (759) & 64 & 55.34\arcsec, 22.50\arcsec, 20.32$^{\circ}$ & 7178.14\,s\\ \hline

\begin{tabular}[c]{@{}c@{}}Pointing\,17\\ 2024 Nov 21 \end{tabular} & \begin{tabular}[c]{@{}c@{}}05$^{\rm h}$21$^{\rm m}$36\fs59\\ $-$67\textdegree{}57\arcmin53\farcs50\end{tabular} & 44 (765) & 58 & 73.83\arcsec, 20.19\arcsec, $-$36.33$^{\circ}$ & 7164.97\,s\\ \hline

\begin{tabular}[c]{@{}c@{}}Pointing\,18\\ 2024 Nov 21 \end{tabular} & \begin{tabular}[c]{@{}c@{}}04$^{\rm h}$51$^{\rm m}$45\fs21\\ $-$67\textdegree{}00\arcmin20\farcs90\end{tabular} & 44 (754) & 58 & 44.57\arcsec, 20.20\arcsec, $-$4.63$^{\circ}$ & 7178.14\,s\\ \hline

\begin{tabular}[c]{@{}c@{}}Pointing\,19\\ 2025 Feb 03 \end{tabular} & \begin{tabular}[c]{@{}c@{}}05$^{\rm h}$36$^{\rm m}$07\fs65\\ $-$70\textdegree{}06\arcmin18\farcs00\end{tabular} & 40 (763) & 58 & 61.54\arcsec, 23.67\arcsec, $-$22.81$^{\circ}$ & 7145.53\,s\\ \hline

\begin{tabular}[c]{@{}c@{}}Pointing\,20\\ 2025 Feb 03 \end{tabular} & \begin{tabular}[c]{@{}c@{}}05$^{\rm h}$54$^{\rm m}$26\fs02\\ $-$68\textdegree{}30\arcmin34\farcs00\end{tabular} & 40 (767) & 58 & 45.44\arcsec, 24.03\arcsec, 0.41$^{\circ}$ & 7180.65\,s\\ \hline

\begin{tabular}[c]{@{}c@{}}Pointing\,21\\ 2025 Mar 16 \end{tabular} & \begin{tabular}[c]{@{}c@{}}04$^{\rm h}$58$^{\rm m}$08\fs10\\ $-$70\textdegree{}23\arcmin05\farcs40\end{tabular} & 44 (766) & 61 & 54.16\arcsec, 21.71\arcsec, $-$15.49$^{\circ}$ & 7160.58\,s\\ \hline

\begin{tabular}[c]{@{}c@{}}Pointing\,22\\ 2025 Mar 16 \end{tabular} & \begin{tabular}[c]{@{}c@{}}05$^{\rm h}$04$^{\rm m}$47\fs21\\ $-$65\textdegree{}19\arcmin52\farcs30\end{tabular} & 44 (762) & 61 & 39.99\arcsec, 21.69\arcsec, 6.96$^{\circ}$ & 7181.28\,s\\ \hline

\end{tabular}
\end{table*}    

\clearpage
\onecolumn

\section{Targeted sources} \label{appendix: Targeted sources}

The following table provides a comprehensive list of all sources targeted in the new TRAPUM LMC Survey pointings presented in this work, comprising 18 MeerKAT pointings (refer to \autoref{section: observations}). For each source, we give the object name, its classification, and the corresponding pointing. The list includes SNRs and candidate SNRs, HMXBs, and GCs, as compiled from the literature and astronomical databases.

\begin{longtable}{lcc}
\caption{List of sources targeted in pointings\,5 to 22, with the addition of the X-ray magnetar candidate found by \protect\cite{Imbrogno2023} in Pointing\,14. The sources were obtained from \protect\cite{Bozzetto2017}, \protect\cite{Maitra2019}, \protect\cite{Yew2021}, \protect\cite{Maitra2021}, \protect\cite{Kavanagh2022}, \protect\cite{Sasaki2022}, \protect\cite{Bozzetto2023}, and \protect\cite{Zangrandi2024} for SNRs and candidate SNRs, and from \protect\cite{Antoniou2016}, \protect\cite{Maitra2019}, \protect\cite{Maitra2021}, and \protect\cite{Haberl2022} for HMXBs. The GCs were obtained from the SIMBAD astronomical database \protect\citep{Wenger2000}.}
\label{tab:sources} \\

\hline
\textbf{Object} & \textbf{Type} & \textbf{Pointing} \\
\hline
\endfirsthead

\multicolumn{3}{c}
{{\tablename\ \thetable{} -- continued from previous page}} \\
\hline
\textbf{Object} & \textbf{Type} & \textbf{Pointing} \\
\hline
\endhead

\hline \multicolumn{3}{r}{{Continued on next page}} \\
\endfoot

\hline
\endlastfoot

MCSNR\,J0537$-$6910 & SNR & \multirow{12}{*}{Pointing\,5} \\
MCSNR\,J0540$-$6920 & SNR & \\
MCSNR\,J0543$-$6858 & SNR & \\
MCSNR\,J0542$-$6852 & Candidate SNR & \\
MCSNR\,J0543$-$6906 & Candidate SNR & \\
MCSNR\,J0543$-$6923 & Candidate SNR & \\
MCSNR\,J0543$-$6928 & Candidate SNR & \\
NGC\,2108 & GC & \\
LXP4.42 & HMXB & \\
RX\,J0541.4$-$6936 & HMXB & \\
RXJ0546.8$-$6851 & HMXB & \\
PSR\,J0540$-$69 & PSR & \\
PSR\,J0543$-$6851 & PSR & \\
\hline

MCSNR\,J0525$-$6559 & SNR & \multirow{14}{*}{Pointing\,6} \\
MCSNR\,J0526$-$6605 & SNR & \\
MCSNR\,J0527$-$6550 & SNR & \\
MCSNR\,J0525$-$6621 & Candidate SNR & \\
MCSNR\,J0532$-$6554 & Candidate SNR & \\
NGC\,1978 & GC & \\
RX\,J0527.3$-$6552 & HMXB & \\
LXP69.2 & HMXB & \\
LXP272 & HMXB & \\
RX\,J0530.7$-$6606 & HMXB & \\
RX\,J0531.2$-$6607 & HMXB & \\
XMMU\,J053118.2$-$660730 & HMXB & \\
RX\,J0532.5$-$6551 & HMXB & \\
LMC\,X-4 & HMXB & \\
\hline

MCSNR\,J0527$-$6714 & SNR & \multirow{15}{*}{Pointings\,7 and 12} \\
MCSNR\,J0528$-$6727 & SNR & \\
MCSNR\,J0529$-$6653 & SNR & \\
MCSNR\,J0532$-$6732 & SNR & \\
MCSNR\,J0536$-$6735 & SNR & \\
MCSNR\,J0528$-$6719 & Candidate SNR & \\
MCSNR\,J0534$-$6700 & Candidate SNR & \\
MCSNR\,J0534$-$6720 & Candidate SNR & \\
LXP28.8 & HMXB & \\
RX\,J0535.0$-$6700 & HMXB & \\
1A0535$-$668 & HMXB & \\
PSR\,J0529$-$6652 & PSR & \\
PSR\,J0532$-$6639 & PSR & \\
PSR\,J0534$-$6703 & PSR & \\
PSR\,J0535$-$66 & PSR & \\
\hline

MCSNR\,J0547$-$6941 & SNR & \multirow{7}{*}{Pointing\,8} \\
MCSNR\,J0547$-$6943 & SNR & \\
MCSNR\,J0543$-$6923 & Candidate SNR & \\
MCSNR\,J0543$-$6928 & Candidate SNR & \\
MCSNR\,J0548$-$6941 & Candidate SNR & \\
MCSNR\,J0549$-$7001 & Candidate SNR & \\
RX\,J0541.4$-$6936 & HMXB & \\
\hline

MCSNR\,J0527$-$7104 & SNR & \multirow{8}{*}{Pointing\,9} \\
MCSNR\,J0531$-$7100 & SNR & \\
MCSNR\,J0534$-$7033 & SNR & \\
MCSNR\,J0536$-$7039 & SNR & \\
NGC\,1987 & GC & \\
NGC\,2031 & GC & \\
XMMU\,J053115.4$-$705350 & HMXB & \\
RX\,J0532.3$-$7107 & HMXB & \\
\hline

MCSNR\,J0454$-$6626 & SNR & \multirow{7}{*}{Pointing\,10} \\
NGC\,1783 & GC & \\
NGC\,1805 & GC & \\
NGC\,1818 & GC & \\
RX\,J0457.2$-$6612 & HMXB & \\
LXP4.10 & HMXB & \\
PSR\,J0502$-$6617 & PSR & \\
\hline

MCSNR\,J0509$-$6731 & SNR & \multirow{7}{*}{Pointing\,11} \\
MCSNR\,J0510$-$6708 & SNR & \\
MCSNR\,J0512$-$6707 & SNR & \\
MCSNR\,J0512$-$6716 & SNR & \\
MCSNR\,J0513$-$6724 & SNR & \\
MCSNR\,J0513$-$6731 & Candidate SNR & \\
MCSNR\,J0513$-$6724 & HMXB & \\
\hline

MCSNR\,J0453$-$6829 & SNR & \multirow{8}{*}{Pointing\,13} \\
MCSNR\,J0455$-$6839 & SNR & \\
MCSNR\,J0450$-$6818 & Candidate SNR & \\
MCSNR\,J0457$-$6823 & Candidate SNR & \\
MCSNR\,J0455$-$6830 & Candidate SNR & \\
MCSNR\,J0456$-$6830 & Candidate SNR & \\
NGC\,1696 & GC & \\
RX\,J0456.9$-$6824 & HMXB & \\
\hline

MCSNR\,J0447$-$6918 & SNR & \multirow{15}{*}{Pointing\,14} \\
MCSNR\,J0449$-$6903 & SNR & \\
MCSNR\,J0449$-$6920 & SNR & \\
MCSNR\,J0456$-$6950 & SNR & \\
MCSNR\,J0451$-$6906 & Candidate SNR & \\
MCSNR\,J0451$-$6951 & Candidate SNR & \\
MCSNR\,J0457$-$6923 & Candidate SNR & \\
NGC\,1756 & GC & \\
LXP187 & HMXB & \\
XMMU\,J045315.1$-$693242 & HMXB & \\
XMMU\,J045736.9$-$692727 & HMXB & \\
4XMM\,J045626.3$-$694723 & Magnetar candidate & \\
PSR\,J0455$-$6951 & PSR & \\
PSR\,J0456$-$69 & PSR & \\
PSR\,J0457$-$69 & PSR & \\
\hline

MCSNR\,J0504$-$6723 & SNR & \multirow{8}{*}{Pointing\,15} \\
MCSNR\,J0505$-$6753 & SNR & \\
MCSNR\,J0505$-$6802 & SNR & \\
MCSNR\,J0457$-$6739 & Candidate SNR & \\
MCSNR\,J0459$-$6757 & Candidate SNR & \\
MCSNR\,J0502$-$6739 & Candidate SNR & \\
NGC\,1806 & GC & \\
PSR\,J0458$-$67 & PSR & \\
\hline

MCSNR\,J0525$-$6938 & SNR & \multirow{8}{*}{Pointing\,16} \\
MCSNR\,J0527$-$6912 & SNR & \\
MCSNR\,J0529$-$7004 & SNR & \\
MCSNR\,J0521$-$6936 & Candidate SNR & \\
RX\,J0523.2$-$7004 & HMXB & \\
RX\,J0527.1$-$7005 & HMXB & \\
RX\,J0529.4$-$6952 & HMXB & \\
PSR\,J0532$-$69 & PSR & \\
\hline

MCSNR\,J0517$-$6759 & SNR & \multirow{5}{*}{Pointing\,17} \\
MCSNR\,J0522$-$6740 & SNR & \\
MCSNR\,J0523$-$6753 & SNR & \\
MCSNR\,J0517$-$6757 & Candidate SNR & \\
MCSNR\,J0523$-$6804 & Candidate SNR & \\
\hline

MCSNR\,J0448$-$6700 & SNR & \multirow{5}{*}{Pointing\,18} \\
MCSNR\,J0453$-$6655 & SNR & \\
MCSNR\,J0454$-$6713 & SNR & \\
MCSNR\,J0451$-$6717 & Candidate SNR & \\
MCSNR\,J0452$-$6638 & Candidate SNR & \\
\hline

MCSNR\,J0530$-$7008 & SNR & \multirow{10}{*}{Pointing\,19} \\
MCSNR\,J0534$-$6955 & SNR & \\
MCSNR\,J0534$-$7033 & SNR & \\
MCSNR\,J0536$-$7039 & SNR & \\
MCSNR\,J0540$-$6944 & SNR & \\
MCSNR\,J0538$-$7004 & Candidate SNR & \\
MCSNR\,J0539$-$7001 & Candidate SNR & \\
LMC\,X-1 & HMXB & \\
PSR\,J0532$-$69 & PSR & \\
PSR\,J0535$-$6935 & PSR & \\
\hline

MCSNR\,J0550$-$6823 & SNR & \multirow{3}{*}{Pointing\,20} \\
NGC\,2156 & GC & \\
NGC\,2159 & GC & \\
\hline

MCSNR\,J0454$-$7003 & SNR & \multirow{7}{*}{Pointing\,21} \\
MCSNR\,J0459$-$7008 & SNR & \\
MCSNR\,J0459$-$7008b & Candidate SNR & \\
IGR\,J05007$-$7047 & HMXB & \\
RX\,J0501.6$-$7034 & HMXB & \\
Swift\,J045558.9$-$702001 & HMXB & \\
PSR\,J0456$-$7031 & PSR & \\
\hline

MCSNR\,J0506$-$6541 & SNR & \multirow{4}{*}{Pointing\,22} \\
MCSNR\,J0506$-$6509 & Candidate SNR & \\
MCSNR\,J0500$-$6512 & Candidate SNR & \\
NGC\,1831 & GC \\
\label{tab: sources}
\end{longtable}
\twocolumn 
\onecolumn

\section{Beam maps}\label{appendix: beam maps}

The following figure shows the beam maps for the 18 MeerKAT pointings presented in this paper (refer to \autoref{section: observations}). The full TRAPUM LMC Survey coverage, including both previously published and additional pointings, will be presented in a future paper.

\begin{figure*}
\centering

\begin{subfigure}{0.46\linewidth}
\includegraphics[width=\linewidth]{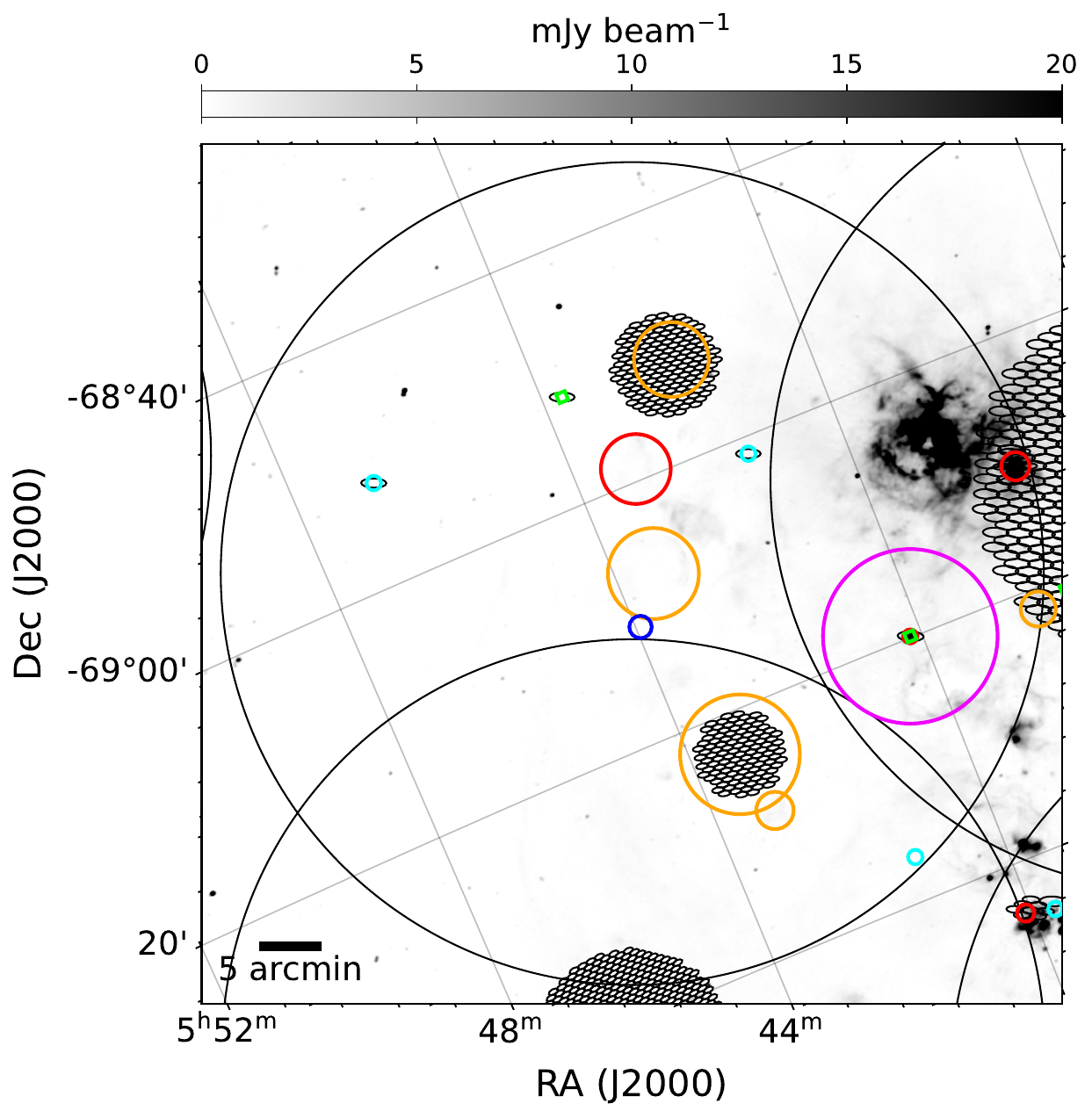}
\caption{Pointing\,5}
\end{subfigure}
\hfill
\begin{subfigure}{0.46\linewidth}
\includegraphics[width=\linewidth]{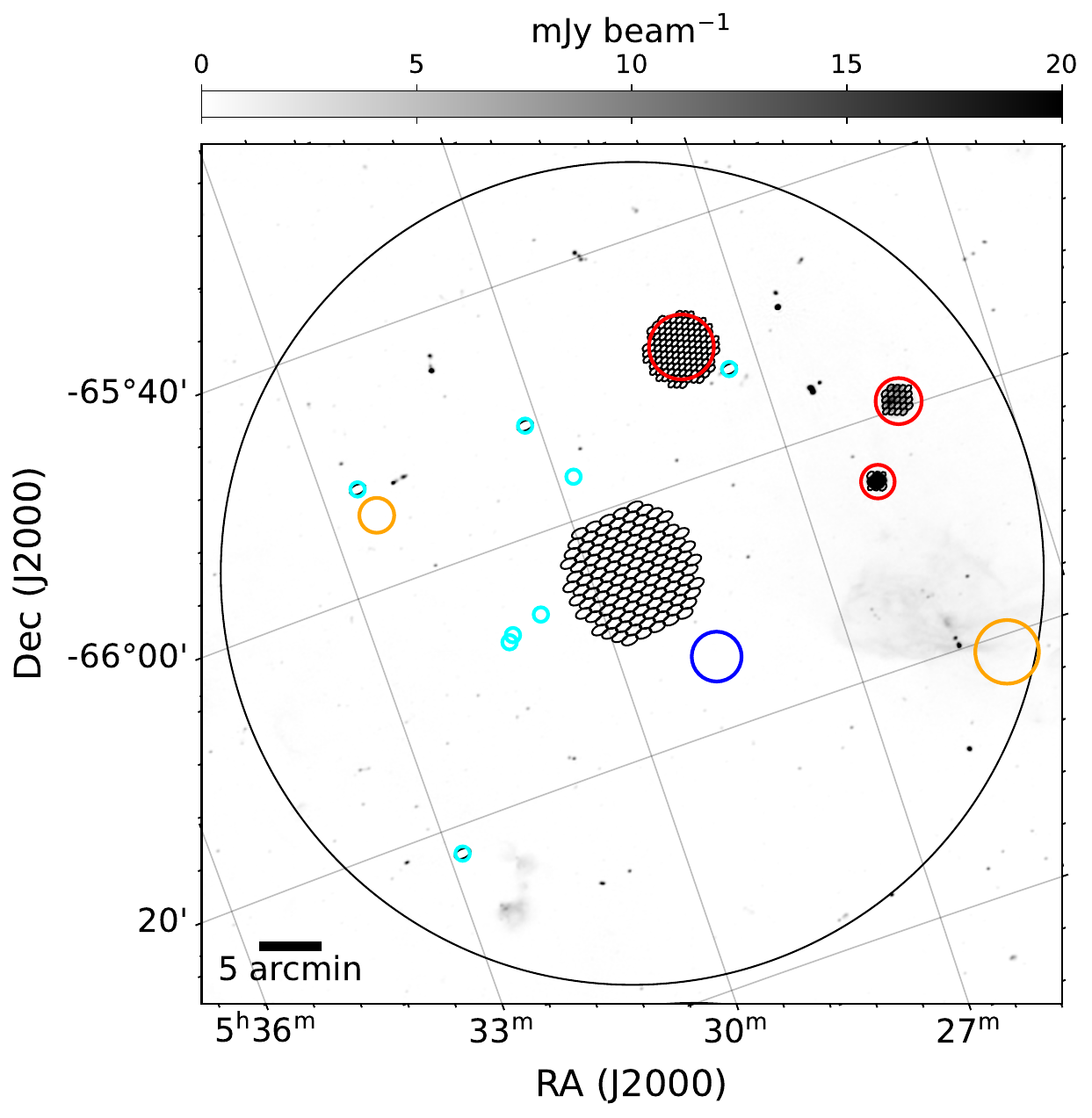}
\caption{Pointing\,6}
\end{subfigure}

\begin{subfigure}{0.46\linewidth}
\includegraphics[width=\linewidth]{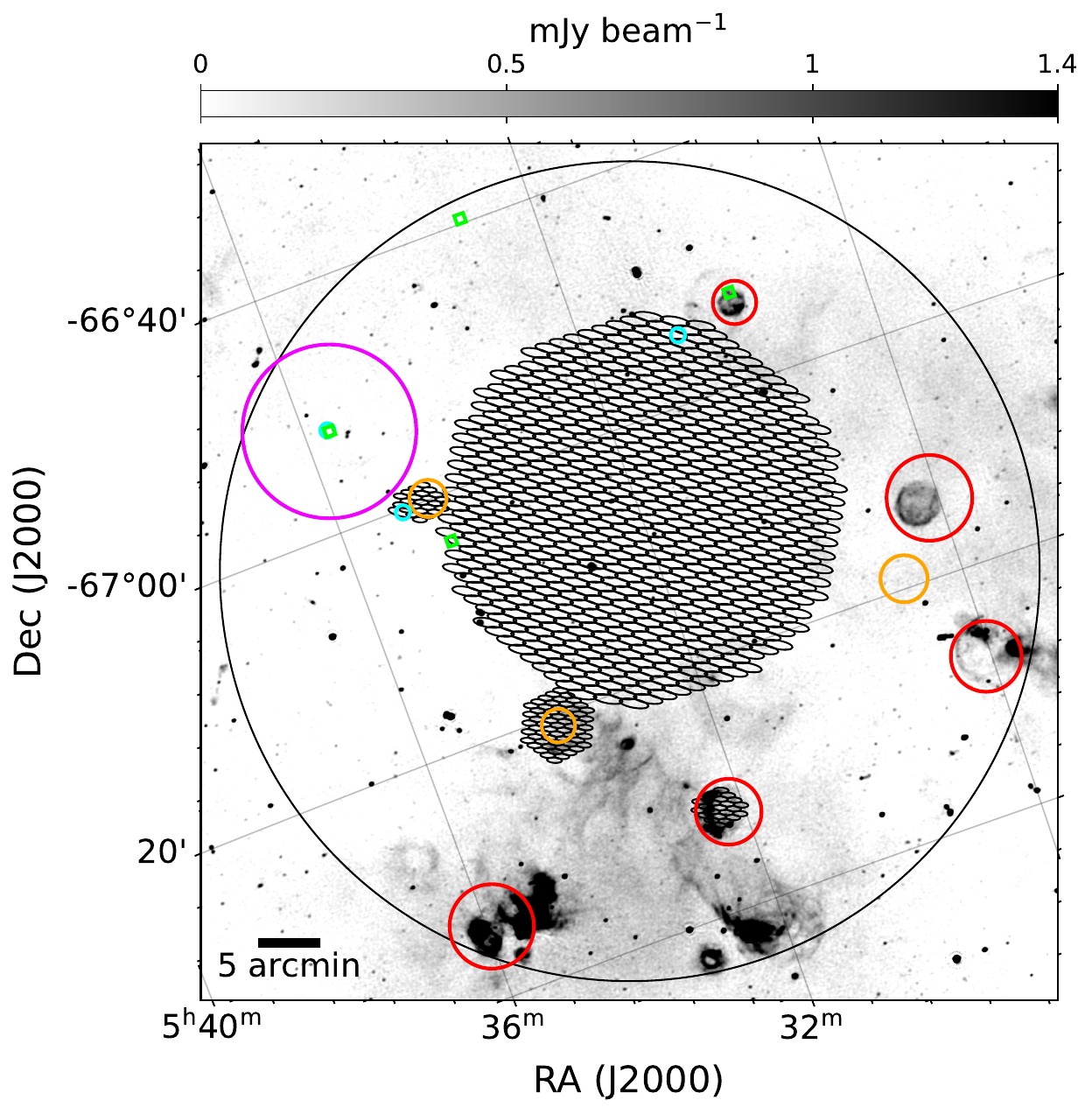}
\caption{Pointing\,7 (re-observed as Pointing\,12)}
\end{subfigure}
\hfill
\begin{subfigure}{0.46\linewidth}
\includegraphics[width=\linewidth]{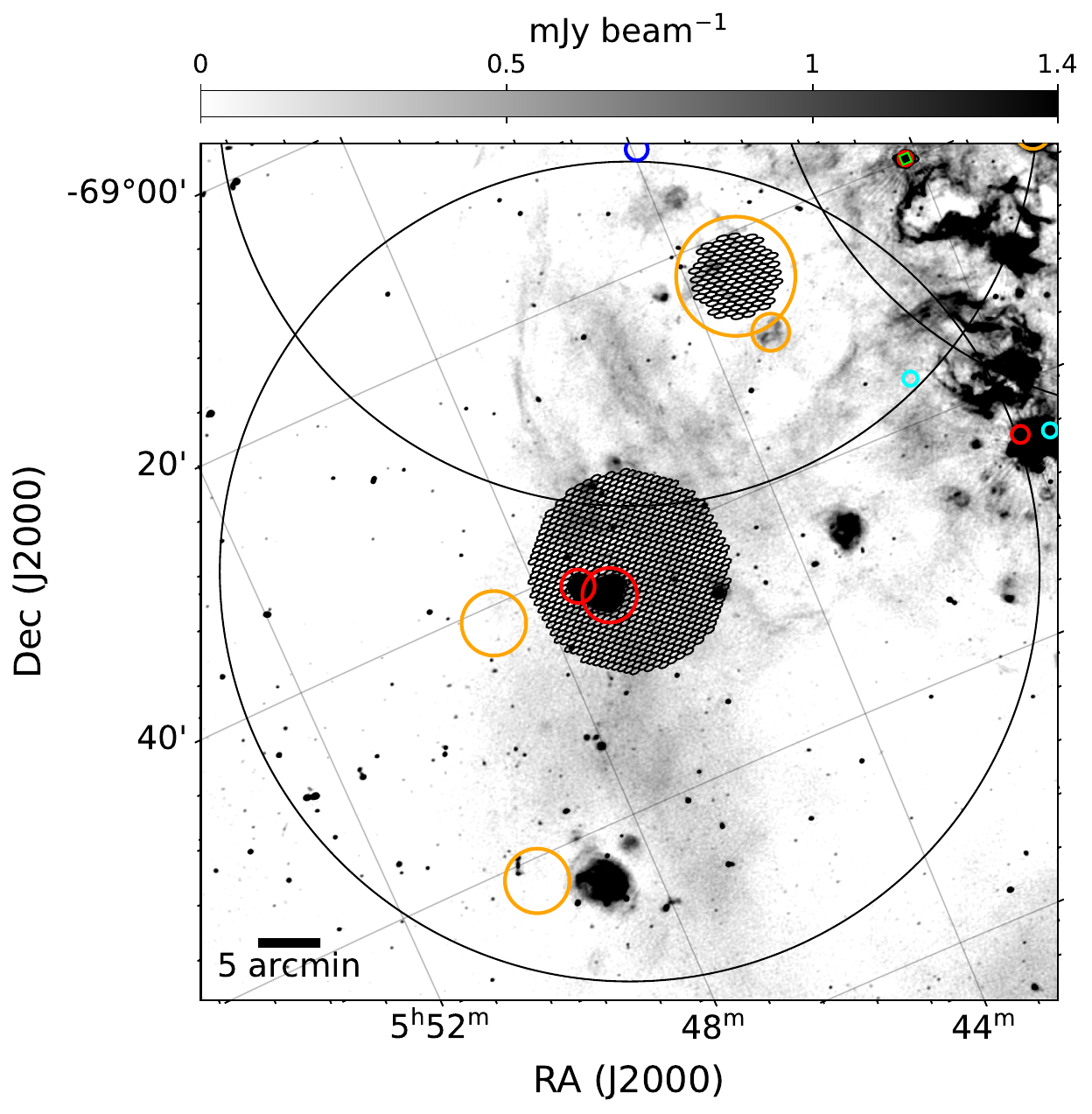}
\caption{Pointing\,8}
\end{subfigure}

\caption{Beam maps for all TRAPUM LMC Survey pointings 5--22, generated using the Python package \textsc{APLpy} \protect\citep{Robitaille2012} with the ASKAP 888\,MHz radio continuum survey image from \protect\cite{Pennock2021} as a background. The outer black circle marks the extent of the MeerKAT incoherent beam at the half-power beam width (HPBW) for the L-band central frequency, 1284\,MHz. Central coherent beams are shown in black and arranged with 50 per cent overlap unless otherwise stated, while additional beams placed on sources of interest use either 50 or 70 per cent overlap. The red circles indicate SNRs, orange circles denote candidate SNRs, green squares mark the nominal positions of known pulsars, blue circles correspond to GCs, cyan circles indicate HMXBs, pink circles show the positional uncertainties of known pulsars, and red stars represent the new pulsar discoveries. Marker sizes reflect the angular extents where applicable, with SNR and candidate SNR sizes from \protect\cite{Zangrandi2024} and GC sizes from the SIMBAD astronomical database (see \autoref{appendix: Targeted sources}). Pointing\,7 was re-observed as Pointing\,12 due to data recording issues, and an overlap of 70 per cent was used for the central tiling beams of Pointing\,8. The pink square in Pointing\,14 represents the nominal position of a candidate X-ray magnetar \protect\citep{Imbrogno2023}.}
\label{figure: all_beam_maps}
\end{figure*}

\begin{figure*}
\ContinuedFloat
\centering

\begin{subfigure}{0.46\linewidth}
\includegraphics[width=\linewidth]{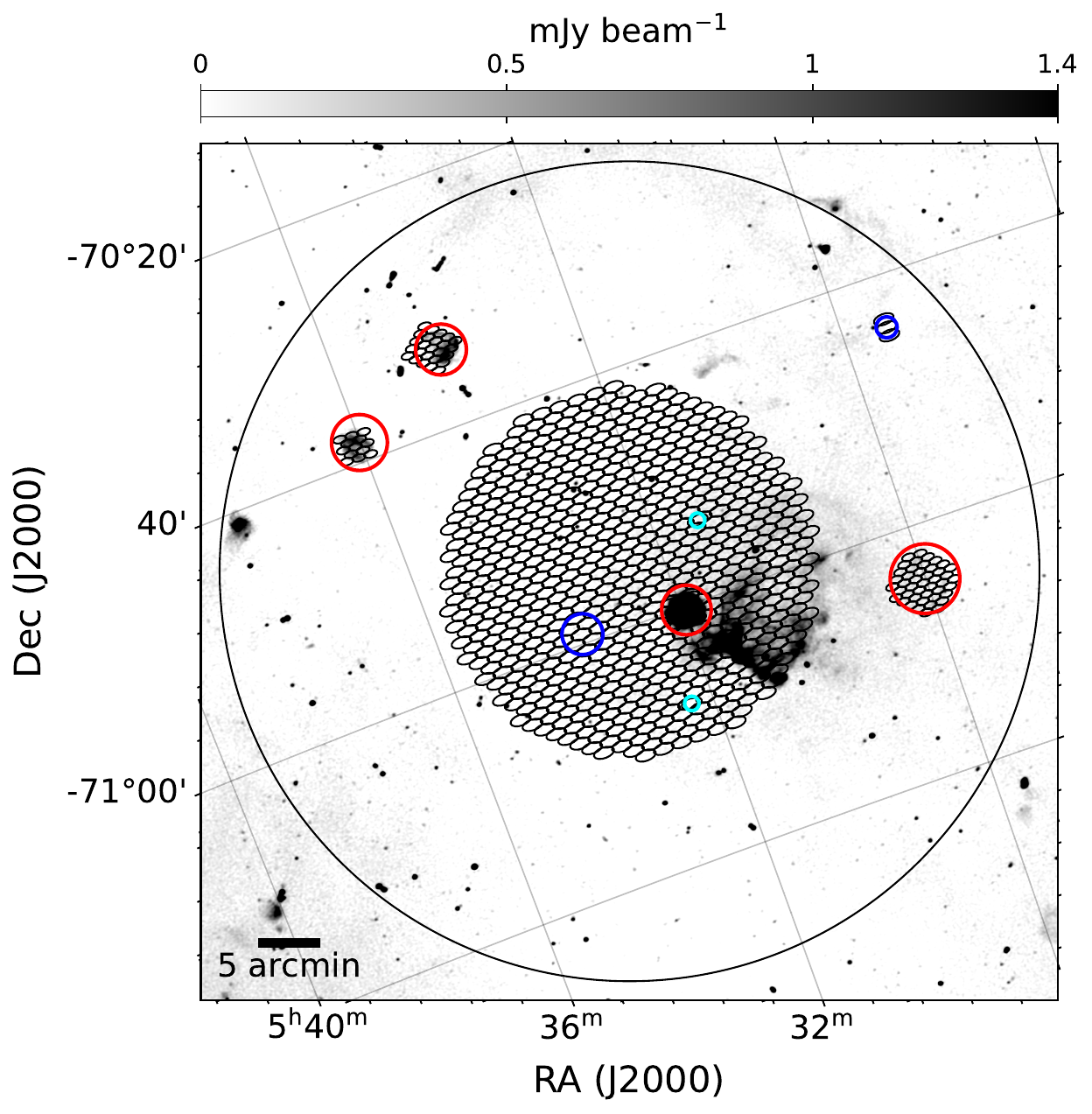}
\caption{Pointing\,9}
\end{subfigure}
\hfill
\begin{subfigure}{0.46\linewidth}
\includegraphics[width=\linewidth]{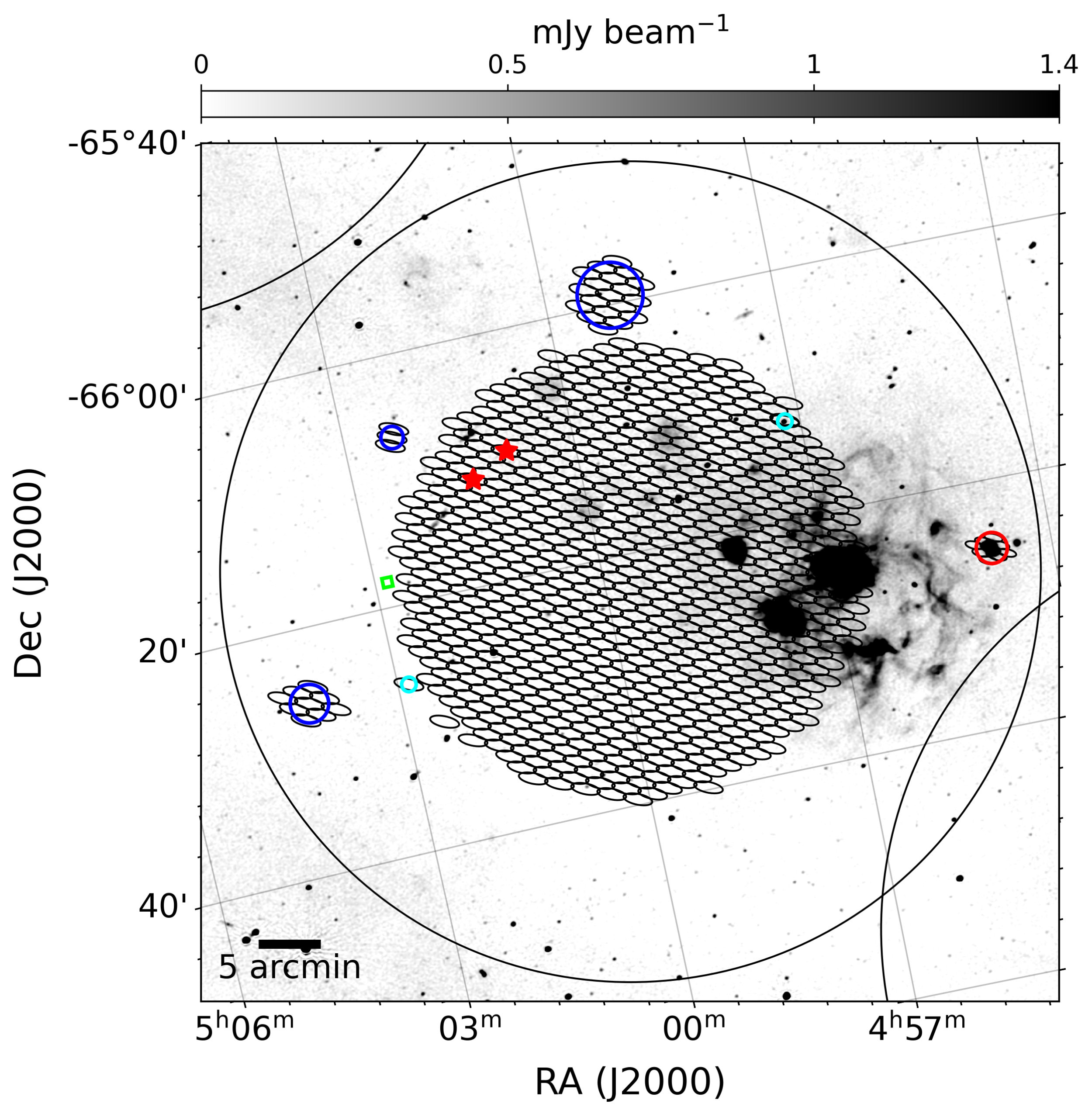}
\caption{Pointing\,10}
\end{subfigure}

\begin{subfigure}{0.46\linewidth}
\includegraphics[width=\linewidth]{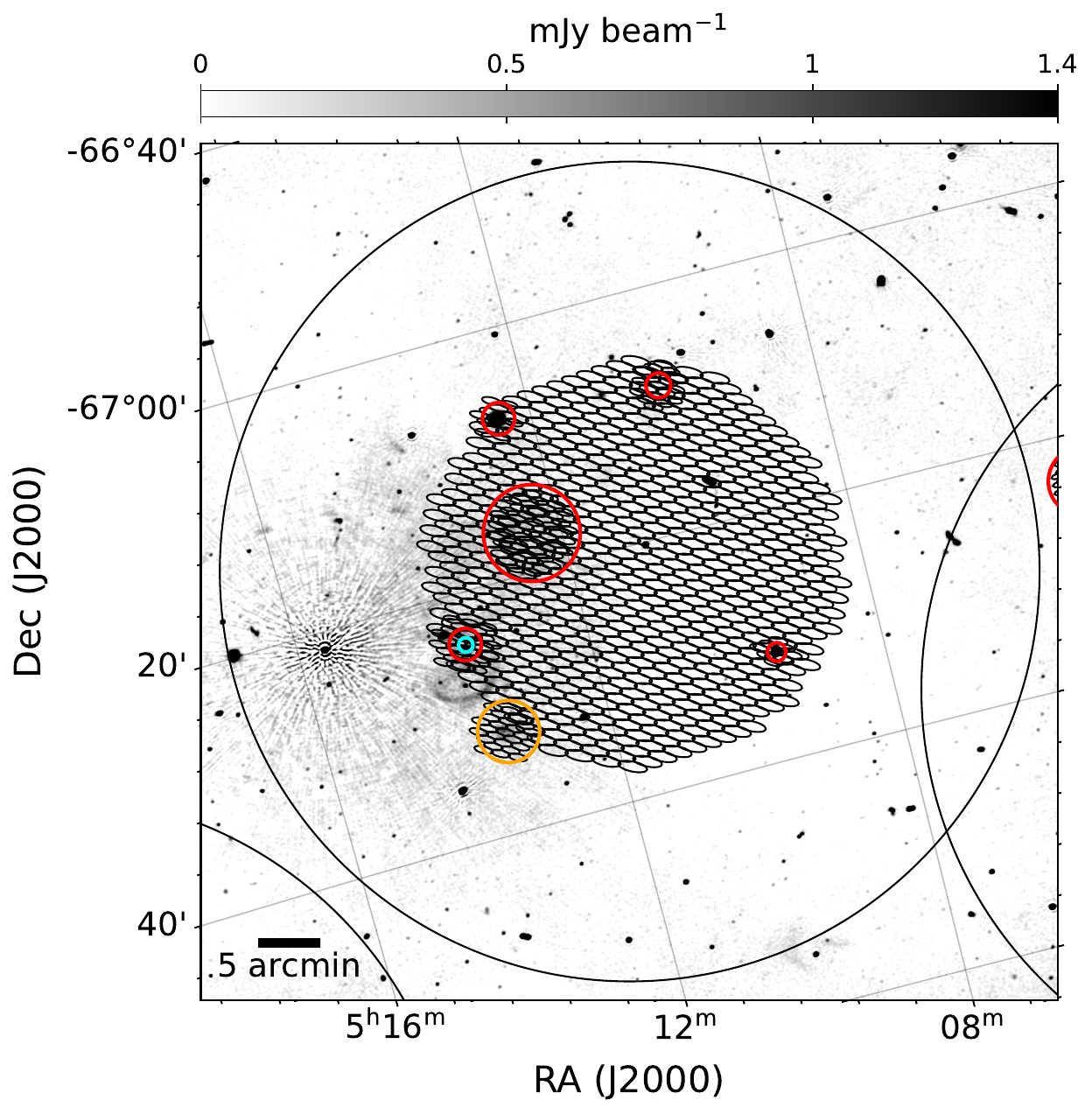}
\caption{Pointing\,11}
\end{subfigure}
\hfill
\begin{subfigure}{0.46\linewidth}
\includegraphics[width=\linewidth]{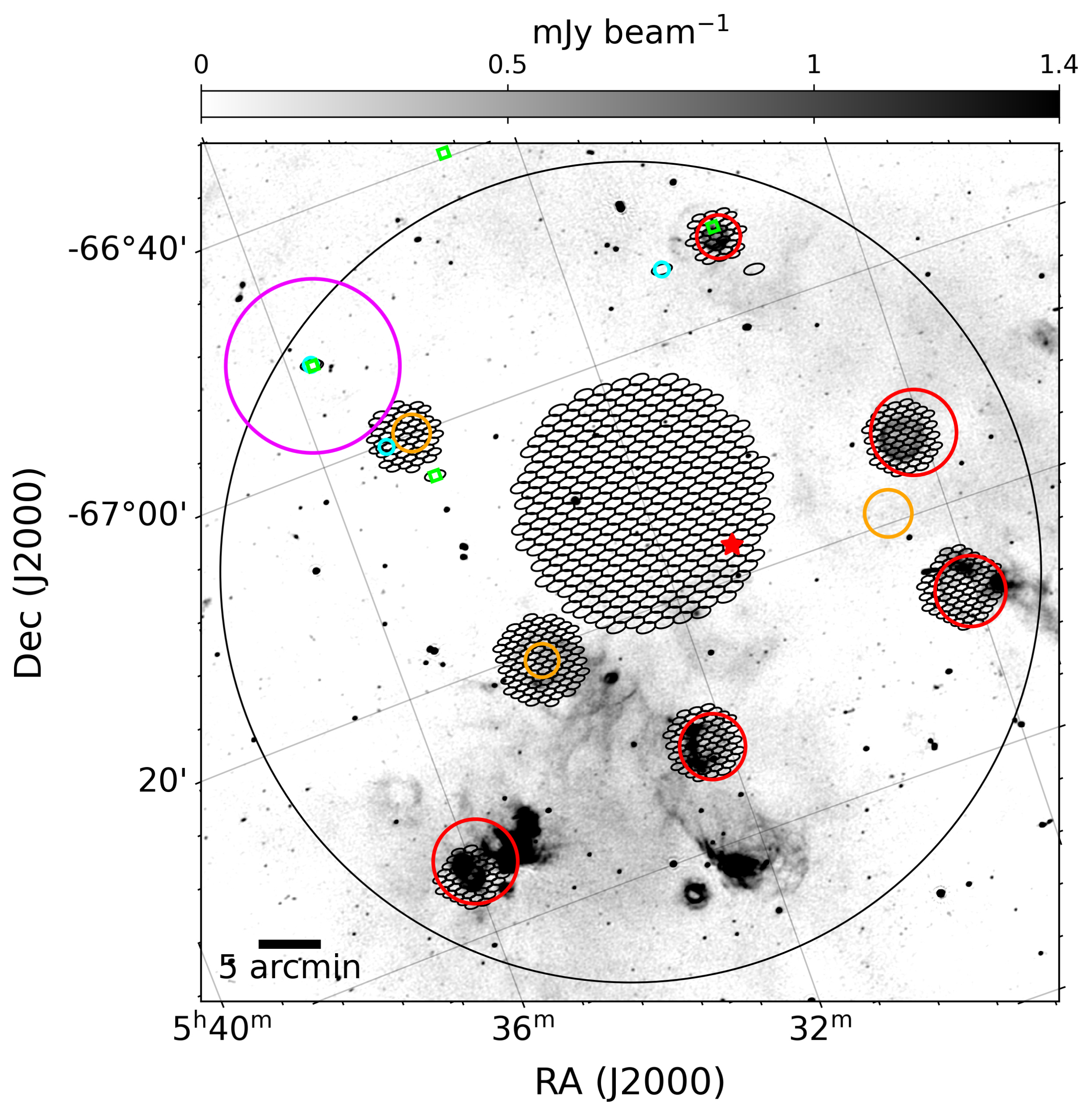}
\caption{Pointing\,12 (re-observation of 7)}
\end{subfigure}

\caption{(Continued)}
\end{figure*}

\begin{figure*}
\ContinuedFloat
\centering

\begin{subfigure}{0.46\linewidth}
\includegraphics[width=\linewidth]{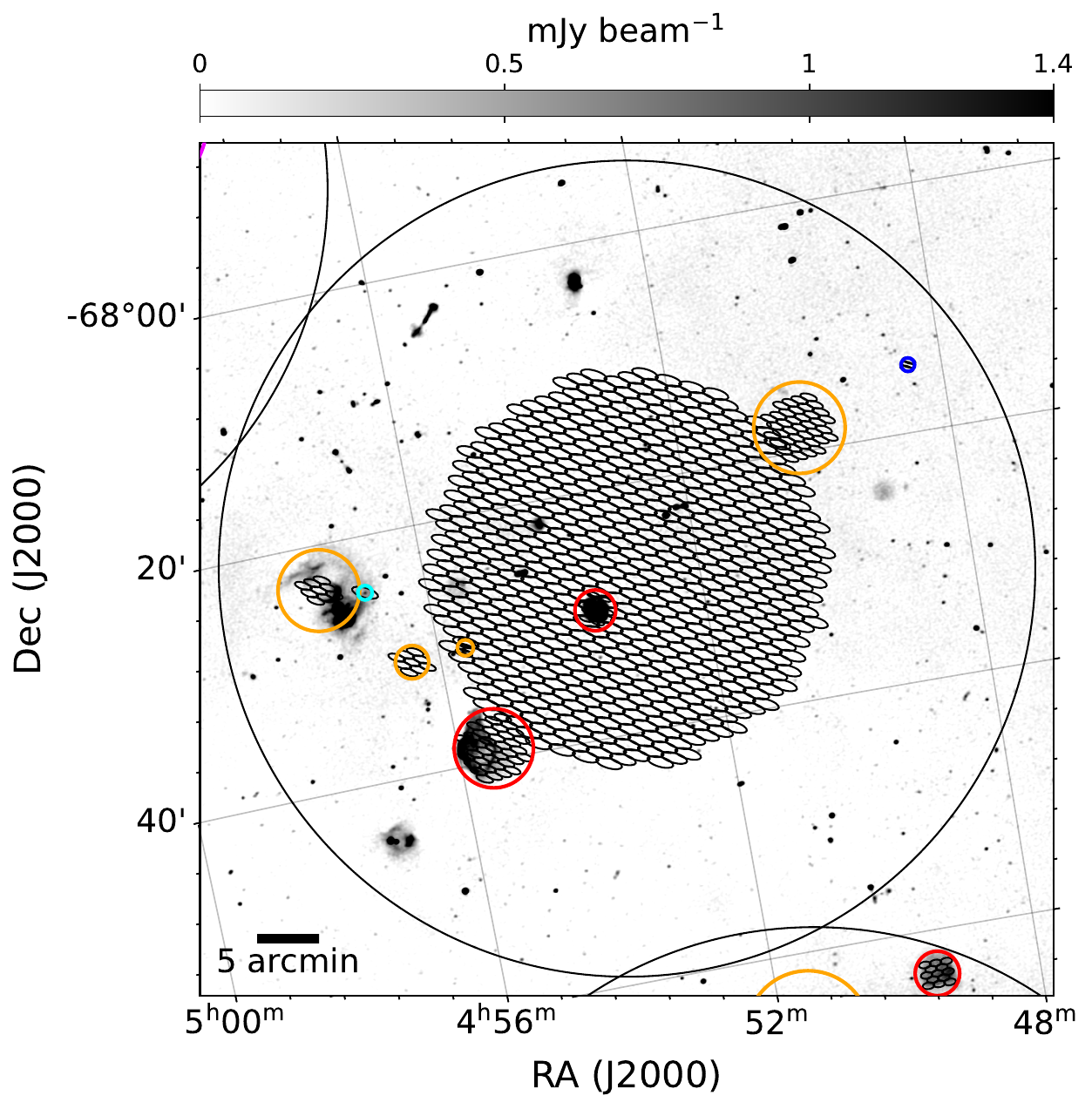}
\caption{Pointing\,13}
\end{subfigure}
\hfill
\begin{subfigure}{0.46\linewidth}
\includegraphics[width=\linewidth]{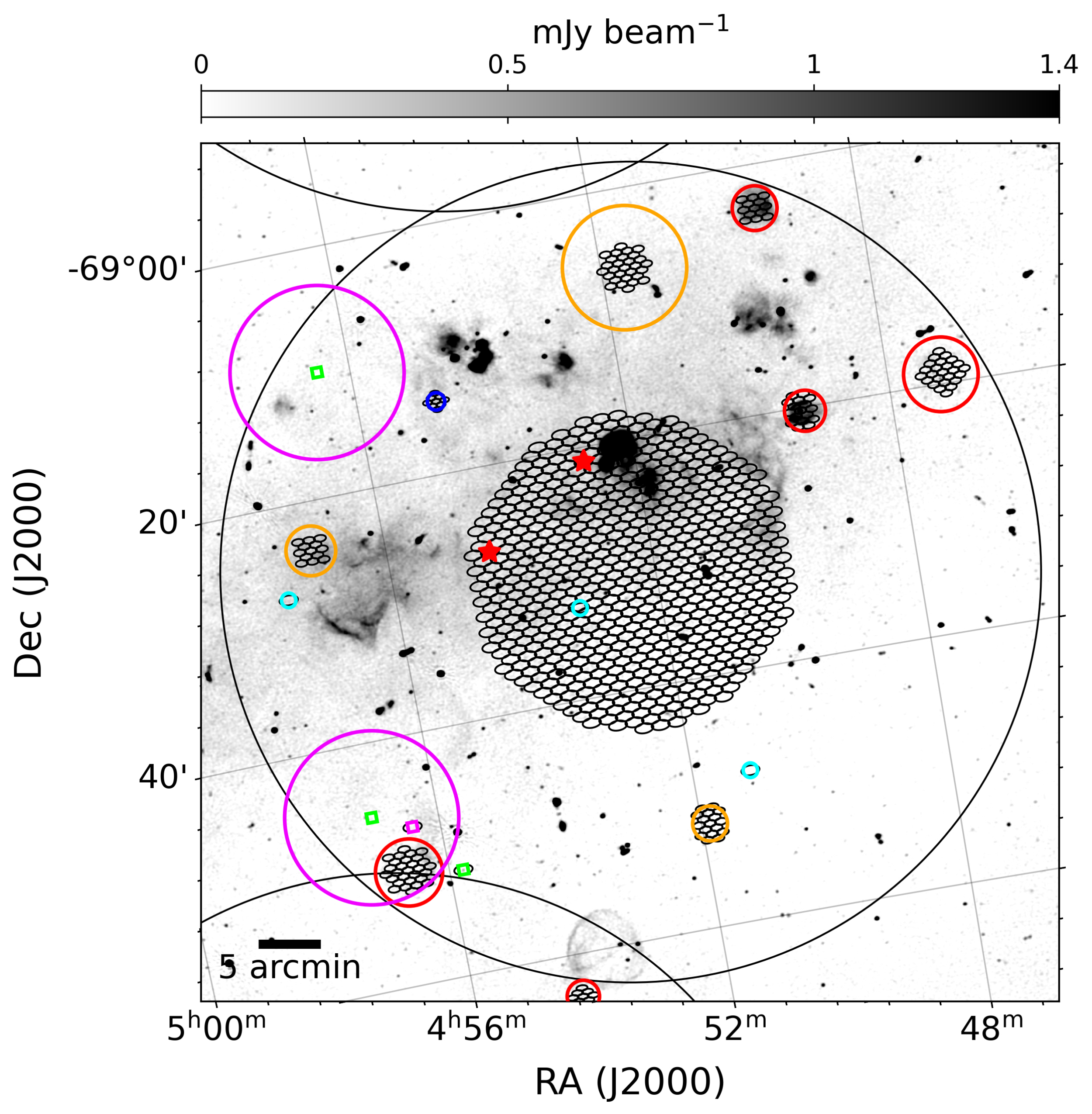}
\caption{Pointing\,14}
\end{subfigure}

\begin{subfigure}{0.46\linewidth}
\includegraphics[width=\linewidth]{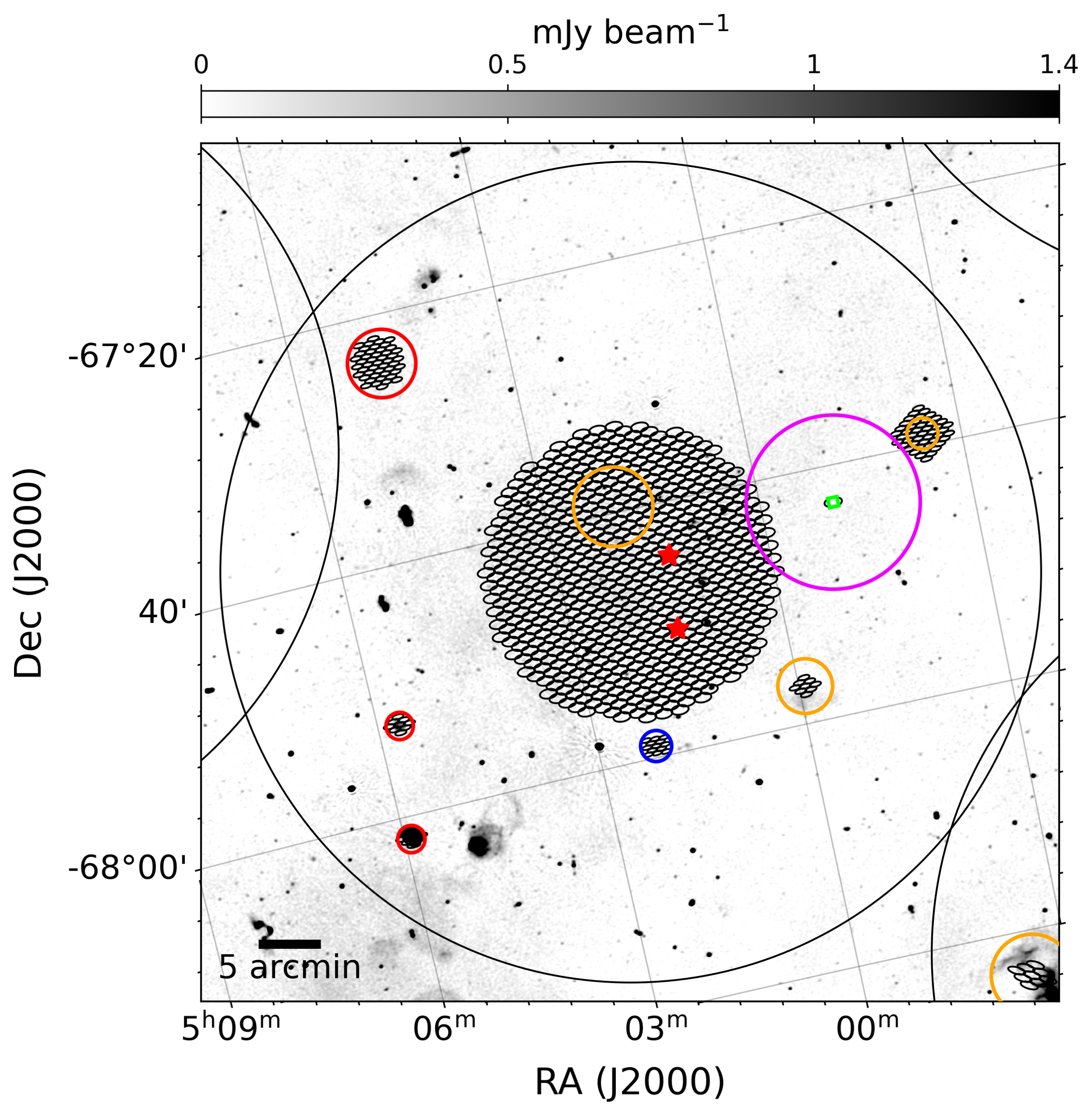}
\caption{Pointing\,15}
\end{subfigure}
\hfill
\begin{subfigure}{0.46\linewidth}
\includegraphics[width=\linewidth]{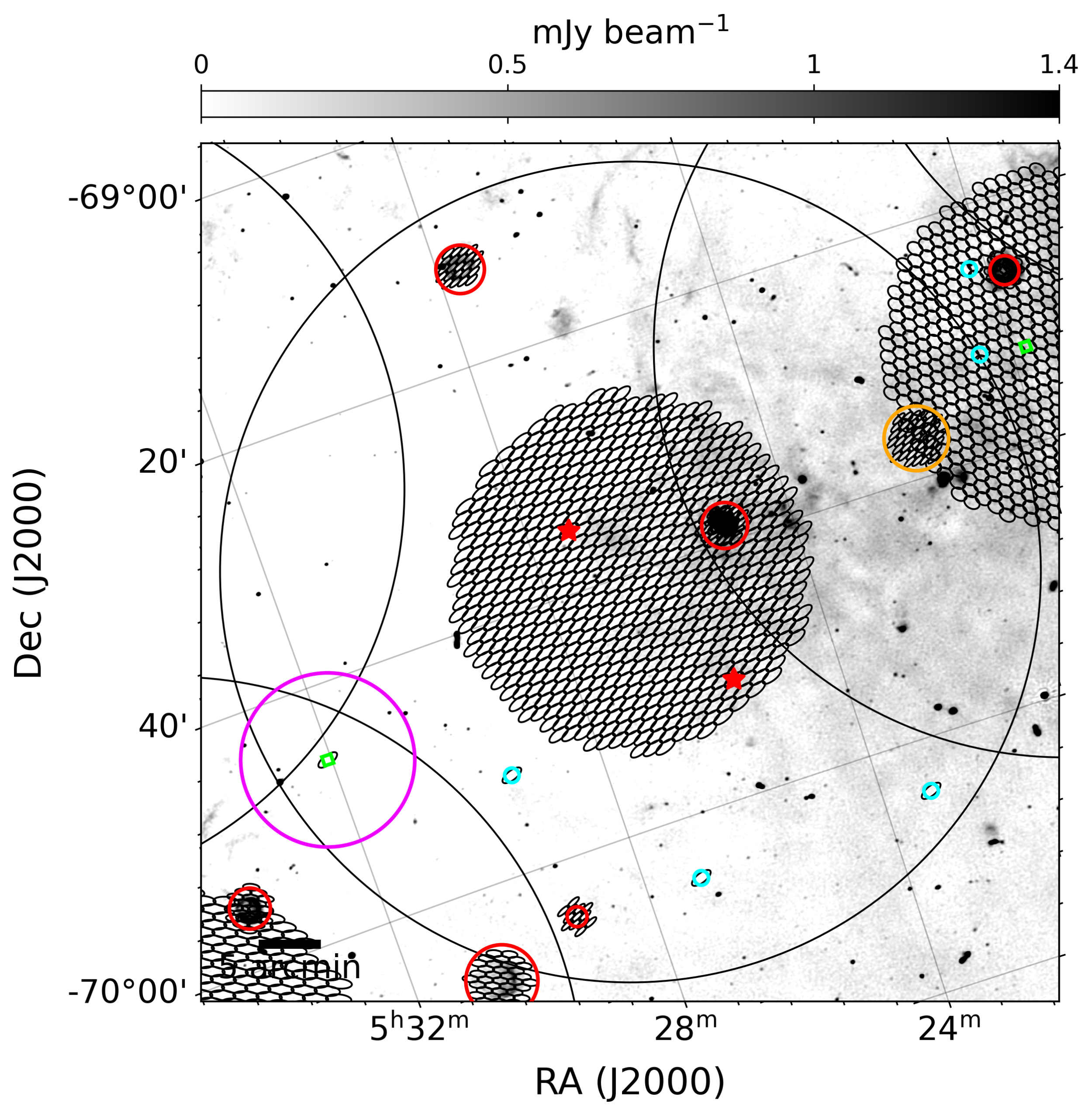}
\caption{Pointing\,16}
\end{subfigure}

\caption{(Continued)}
\end{figure*}

\begin{figure*}
\ContinuedFloat
\centering

\begin{subfigure}{0.46\linewidth}
\includegraphics[width=\linewidth]{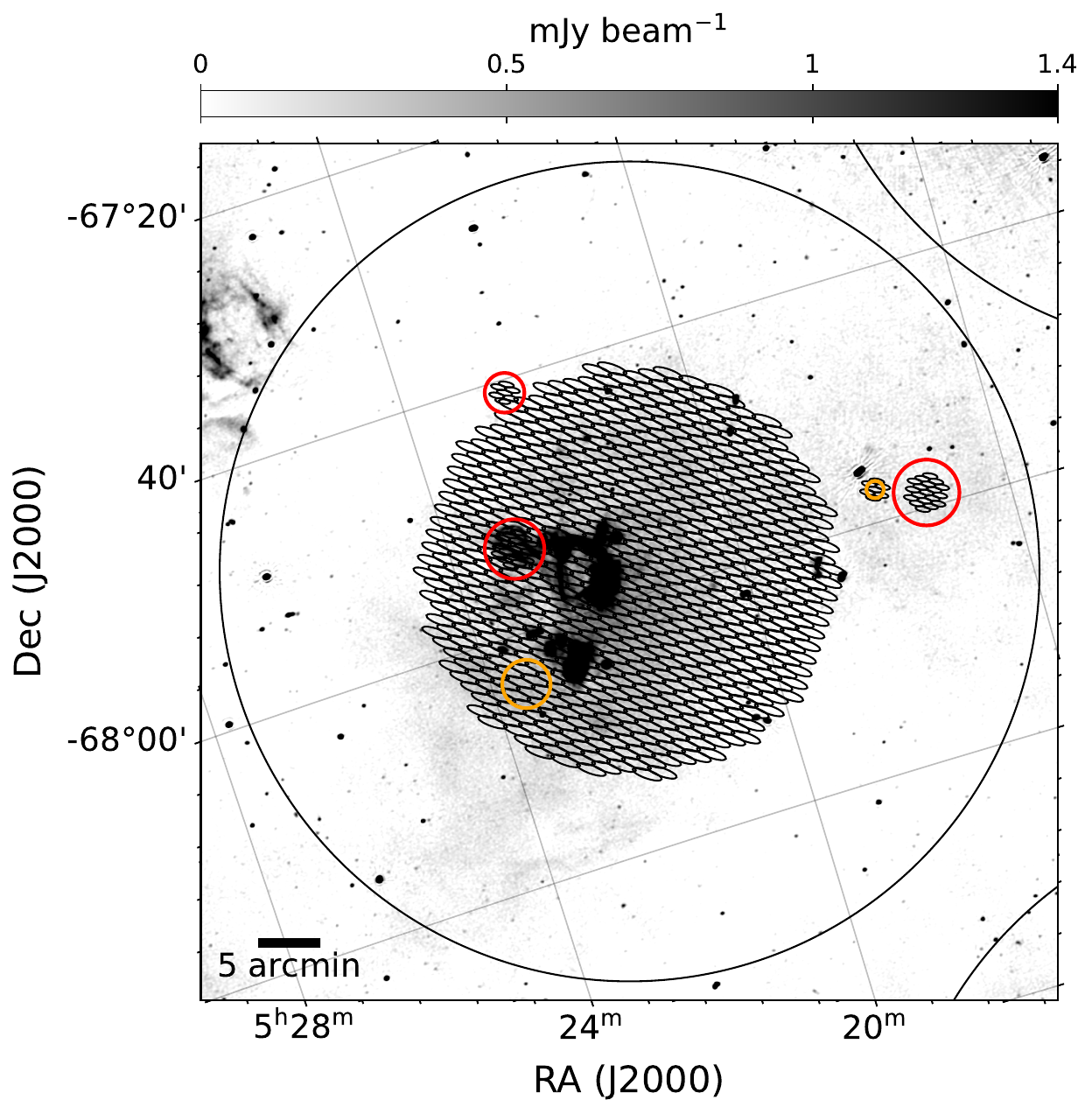}
\caption{Pointing\,17}
\end{subfigure}
\hfill
\begin{subfigure}{0.46\linewidth}
\includegraphics[width=\linewidth]{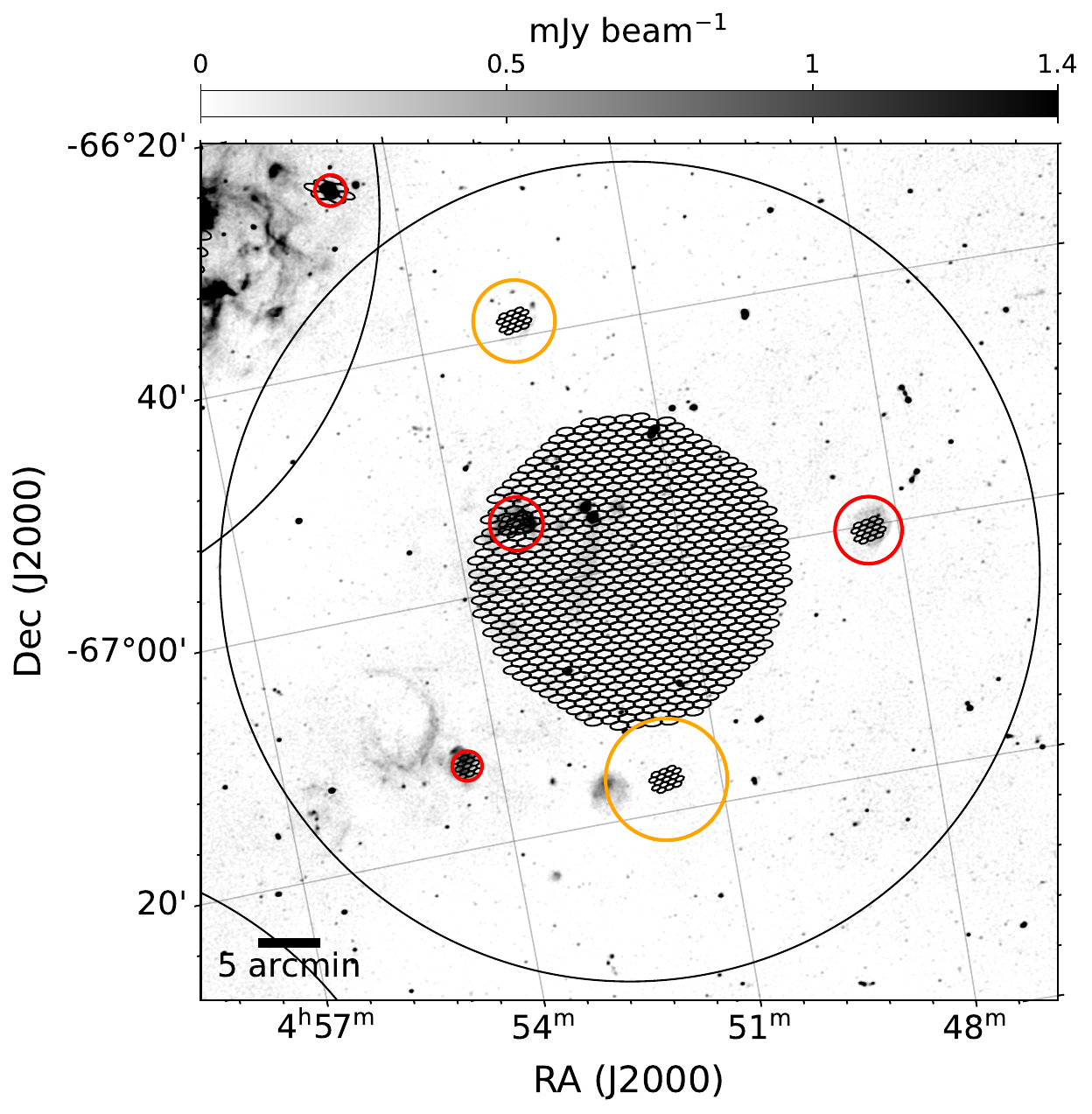}
\caption{Pointing\,18}
\end{subfigure}

\begin{subfigure}{0.46\linewidth}
\includegraphics[width=\linewidth]{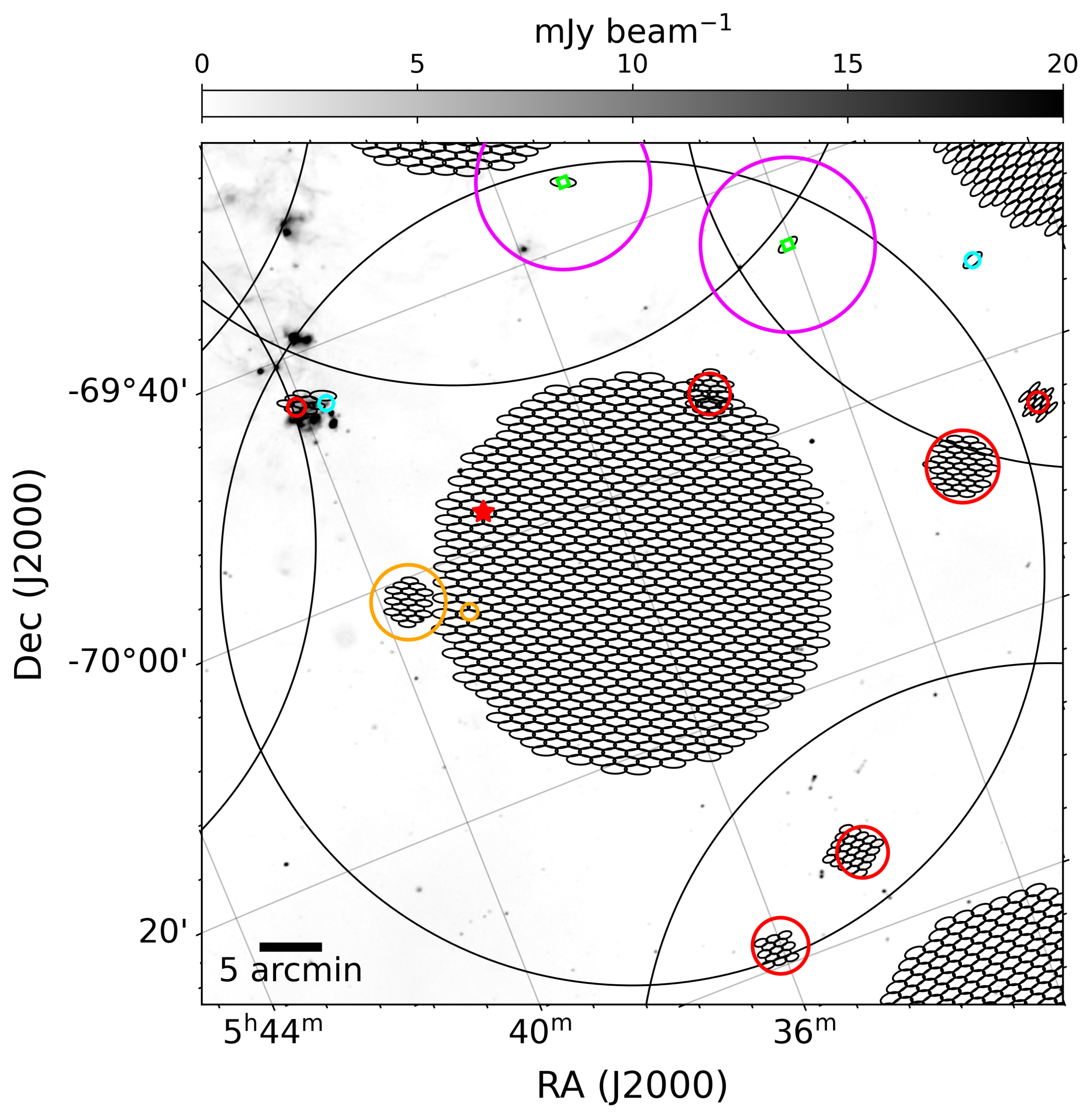}
\caption{Pointing\,19}
\end{subfigure}
\hfill
\begin{subfigure}{0.46\linewidth}
\includegraphics[width=\linewidth]{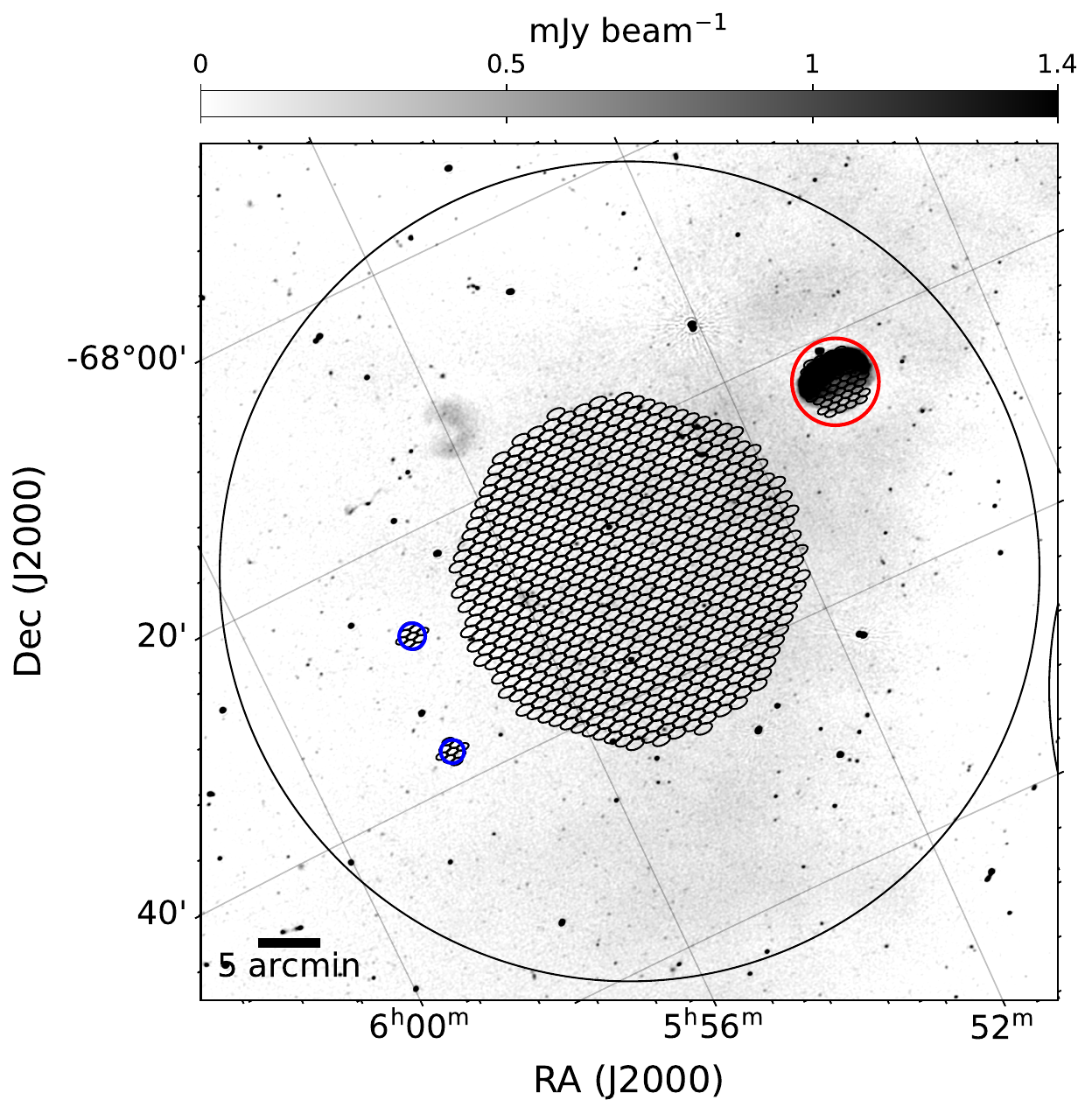}
\caption{Pointing\,20}
\end{subfigure}

\caption{(Continued)}
\end{figure*}

\begin{figure*}
\ContinuedFloat
\centering

\begin{subfigure}{0.46\linewidth}
\includegraphics[width=\linewidth]{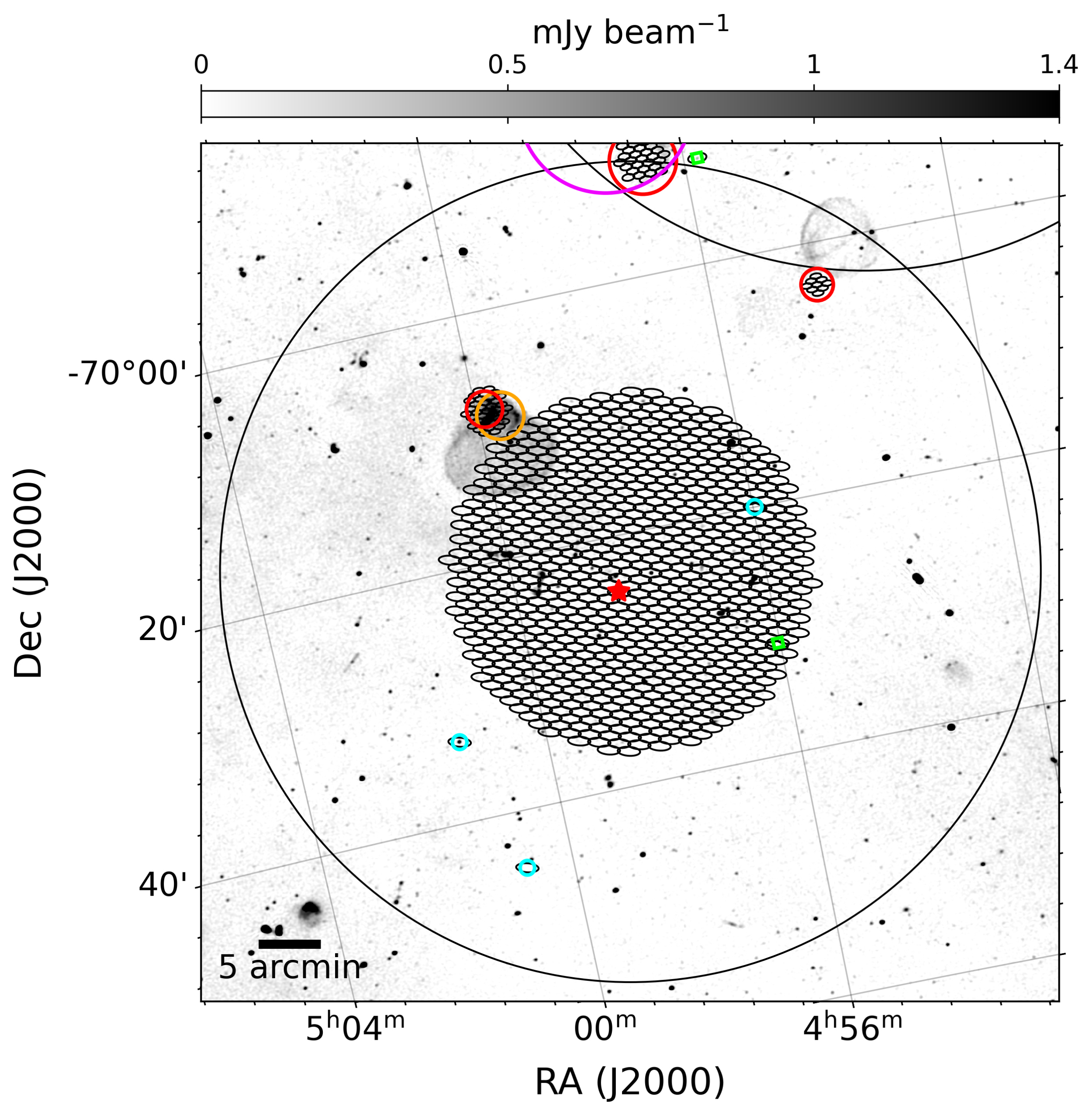}
\caption{Pointing\,21}
\end{subfigure}
\hfill
\begin{subfigure}{0.46\linewidth}
\includegraphics[width=\linewidth]{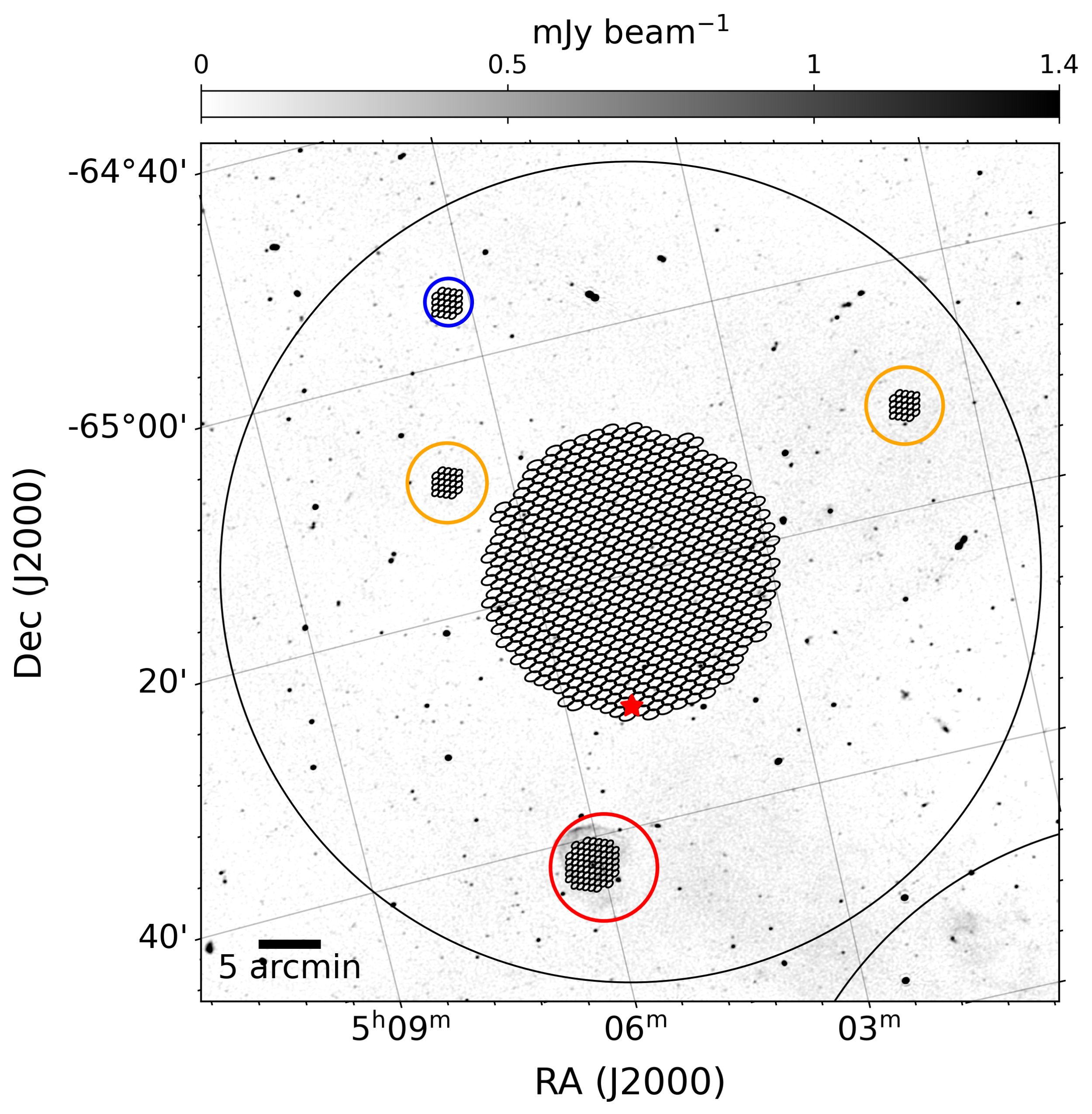}
\caption{Pointing\,22}
\end{subfigure}

\caption{(Continued)}
\end{figure*}

\bsp	
\label{lastpage}
\end{document}